\newcommand{\beq}{\begin{equation}}
\newcommand{\eeq}{\end{equation}}

\documentclass[aps,prd,showpacs]{revtex4}
\usepackage{epsfig}
\usepackage{natbib}
\usepackage{amsmath}
\usepackage{times}
\usepackage{psfrag}

\usepackage{graphicx}
\usepackage{subfigure}
\usepackage{float}
\DeclareMathOperator*{\argmax}{arg\,max}

\begin{document}

\title{Characterization of escape times of Josephson Junctions for signal detection}

\author{Paolo Addesso}
\affiliation{Dept. of Electrical Engineering and Information Engineering, Via Ponte Don Melillo, 1, I-84084 Fisciano, IT}
\author{Giovanni Filatrella}
\affiliation{Dept. of Sciences for Biological, Geological, and Environmental Studies, 
University of Sannio, Via Port'Arsa, 11, I-82100 Benevento, IT}
\author{Vincenzo Pierro}
\affiliation{Dept. of Engineering, University of Sannio, Corso Garibaldi, 107, I-82100 Benevento, IT}


\maketitle

\section*{Abstract}
The measurement of the escape time of a Josephson junction might be used to detect 
the presence of a sinusoidal signal embedded in noise when standard signal processing tools can be prohibitive due to the extreme weakness 
of the source or to the huge amount of data. 
In this paper we show that the prescriptions for the experimental set-up and some physical behaviors depend on the detection strategy.
More specifically, by exploiting the sample mean of escape times to perform detection, two resonant regions are identified. 
At low frequencies there is a stochastic resonance/activation phenomenon, 
while near the plasma frequency a geometric resonance appears. 
Furthermore detection performance in the geometric resonance region is maximized at the prescribed value of the bias current.
The naive sample mean detector is outperformed, in terms of error probability, by the optimal likelihood ratio test.
The latter exhibits only geometric resonance, showing monotonically increasing performance as the bias current approaches 
the junction critical current. 
In this regime the escape times are vanishingly small and therefore performance are essentially limited by measurement electronics.
The behavior of the likelihood ratio  and sample mean detector for different values of incoming signal to noise ratio are 
discussed, and a  relationship with the error probability is found.
The likelihood ratio test based detectors could be employed also to estimate unknown parameters in the applied input signal.
As a prototypical example we study the phase estimation problem of a sinusoidal 
current, that is accomplished  by using the filter bank approach. 
Finally we show that for a physically feasible detector the performances are found to be very close to the Cramer-Rao theoretical bound. Applications might be 
found for example in some astronomical detection problems (where the all sky 
gravitational/radio wave search of pulsar requires the analysis of nearly sinusoidal long lived waveforms at very low signal to noise ratio) 
or to analyze weak signals in the sub-terahertz range (where the traditional electronics counterpart is difficult to implement).

\section{Introduction}
Threshold detection \cite{BulsaraZador, Novotny}
is based on the possibility to ascertain the presence of a signal via the transition from a metastable state to another. 
Thus, the essential ingredients are: i) a two-state detector, ii) a switch from a state to the other induced by the signal, iii) the possibility to detect the transitions when the detector observable crosses a threshold. Under such conditions the original signal is transformed in a series of time-intervals, i.e. the residence times in each state. 

From the physical point of view, the joint action of the applied external signal and the fluctuations induces an escape from a metastable state of the detector. 
The distribution of the escapes acquires, under very general assumptions, the shape of an activation law {\it \'a la Arrhenius}
with an effective energy barrier that depends upon the fluctuations spectrum and the signal (or perturbation) shape \cite{dykman79,graham84,dykman01}. 
The general idea beyond threshold detectors is that, in view of the exponential character of the activation law, residence times are very sensitive to small signals. 
Analyzing escape time series entails a loss of information w.r.t. the direct signal observation. Conversely, it may be useful to shorten the amount of data to be analyzed.
In this perspective, threshold detection has been examined through the lens of information theory \cite{goychuk09} and/or
signal processing \cite{inchbulsa, galdi}.

In this paper we propose to characterize underdamped Josephson Junctions (JJ) as detectors based on the statistics of the escape times. 
It is well known that JJ are superconducting elements that can operate at extremely low temperatures (as low as refrigeration allows) and hence they are affected by a low intrinsic noise. 
Let us remark that to take advantage of the speed and low noise features of JJ it is necessary to analyze escape times, inasmuch the dynamics of the Josephson phase is not directly observable. Instead, the escape from the metastable static solution 
causes a voltage step, associated to the average phase derivative, that is actually detectable. 

Starting from the pioneering papers of Refs. \cite{devoret85,Devoret}, a lot of experimental work has been performed on 
the role of noise in ac driven JJ to detect the phase of the applied signal \cite{yu}, to study  quantum computation \cite{ustinov09} or ratchet effects \cite{carapella01}. 
Another remarkable topic is related to employ JJ as a threshold detector to 
characterize (up to higher order moments) weak fluctuations close to the quantum limit \cite{tobiska04,pekola05,huard07,Barone10} 
(albeit there is also  some controversial discussion about role of quantum noise in JJ \cite{Cirillo}).
In the ac driven quantum regime \cite{levinson} the use of JJ  is particularly appealing for signal detection since  
ways have been devised to minimize the environmental disturbances to the unavoidable quantum level \cite{Lobb}. 

JJ-based detection schemes are very interesting for the applications when the source is so weak that an intermediate amplification stage can introduce too much undesirable noise. Indeed SQUIDs  \cite{BulsaraSQUID} are sensitive to magnetic flux that is a fraction of the quantum flux, $ \Phi_0 \sim 10^{-15}$ Wb. 
The operating frequency range (up to the terahertz timescale) is another appealing feature of JJ. In this connection it can be argued that interesting applications are in the field of terahertz sensing and in the analog processing \cite{mittleman}. 
Another remarkable potential application is related to the search for gravitational wave sources where standard signal processing techniques (i.e. matched filter), essentially based on the intensive use of Fast Fourier Transform algorithm (FFT), 
are computationally prohibitive because applied to a large amount of data (characterized by a very small SNR) \cite{krolak}. We remark that these examples are merely suggestive because our analysis is a proof of principle, not a carefully examination of the applications
and of the technical problems.


It should also be noted that a lot of theoretical problems concerning the proposed dynamical system are currently under study. Indeed, a resonant behavior, that can further improve detection, is expected to occur when the time scale of the external signal matches the fluctuation induced escape time \cite{Benzi81}. Thus JJ are also a playground for topics of active research such as  Stochastic  Resonance  \cite{mcnamara89} or Stochastic Activation \cite{StocAct, StocAct2, StocAct3} that could arise in such a scenario.

The purpose of this paper is to provide a characterization of the JJ as a detector of a periodic signal. The aim is to show that 
nonlinear analogue processing, such as to apply the signal to a JJ, might represent a viable alternative. In fact the signal could be applied and 
processed at very high frequency (for JJ are very fast electronic devices) and with very little extra-noise from the nonlinear device 
(for JJ are superconducting elements that work also at very low temperatures).

It has been proven that a simple statistical analysis of JJ output, such as the Sample Mean (henceforth SM) of transition times, could detect the presence of a sinusoid embedded in thermal noise \cite{filatrella10}.
A more effective analysis of the transition times can be performed by means of appropriate methods of statistical signal processing \cite{helstrom, Shao}, such as the Likelihood Ratio Test (LRT).  
As expected LRT performances depend upon a suitable selection of the physical JJ parameters.
The lack of analytical results for the escape time distribution in the so called underdamped case compels us to use extensive numerical simulations in order to search the JJ optimal working point in a feasible range of the relevant parameters. 

The paper is organized as follows: in Sec. II we briefly describe the physical model of the periodically driven JJ that determines the escape times. In Sec. III we describe the statistical tools used to analyze the escape times, while some technical details are deferred to the Appendices. In Sec. IV we present numerical results that clarify the effectiveness of the statistical analysis, both for signal detection and parameter estimation, namely the sinusoidal phase. Moreover we perform the JJ physical parameters  optimization in order to further improve the detection of a perfectly known sinusoidal signal. Section V,  is devoted to the conclusions.

\section{Escape times distribution: Physical Model and Motivation}

A JJ biased with a sinusoidal signal of amplitude $S_0$ and corrupted by additive noise $\xi(t)$ is modeled by the Langevin equation \cite{BarPat},
\beq
\frac{C \hbar}{2e} \frac{d^2\varphi}{dt^2}+\frac{\hbar }{R 2e} \frac{d\varphi}{dt}+I_c \sin(\varphi)=
I_B+ S_0 \sin(\Omega t+\varphi_0)+ \sqrt{D} \xi(t) + \sqrt{k_b T/R} ~~ n(t).
\label{eq:joseqphu}
\eeq

\noindent Here $C$ and $R$ are the capacitance and the resistance of the JJ, respectively (we consider JJ in the underdamped regime, because the capacitance is not negligible).
Furthermore $I_c$ denotes the Josephson critical current, while $I_B$ is the bias current, $k_b$ denote the Boltzmann constant and $T$ the JJ temperature. The terms $\xi(t)$ $n(t)$ are white Gaussian noise stochastic variables, whose correlators read $< n(t) n(t') > = 2 \delta(t-t')$, $< \xi(t) \xi(t') > = 2 \delta(t-t')$.

\begin{figure}
\centerline{\includegraphics[scale=0.5]{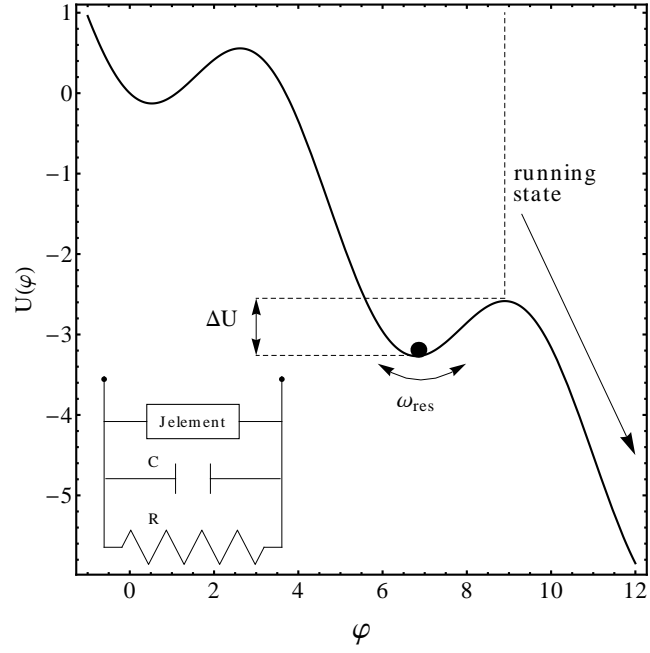}}
\caption{Schematic of the escape process. The junction switches over the energy barrier $\Delta U$ and gives rise to a voltage signal in the running state.
The inset shows the electric circuit model of Eq. (\ref{eq:joseqphu})
}
\label{fig:potential}
\end{figure}

In Eq. (\ref{eq:joseqphu}) thermal fluctuations $\sqrt{k_b T/R}$ can be neglected with respect to the signal noise $D$ if $T\ll D R/k_b$. In fact JJ can be cooled down at
a temperature $T$ much below the signal noise temperature, thus in the following of the paper we will assume that the stochastic 
component is dominated by the signal fluctuations of intensity $D$. The condition to neglect thermal fluctuations is favoured when the junction resistance is high, i.e. when dissipation is low, for Eq. (\ref{eq:joseqphu}) is based on a parallel lumped circuit model (see the inset of Fig. \ref{fig:potential}). In Sec. IV B we will show that low dissipation also favours detection, thus reinforcing the advantages of high $R$.

Introducing the dimensionless time $\tau = \omega_j t$, normalized respect to the characteristic frequency (called Josephson frequency) $\omega_j=[2eI_c/(\hbar C)]^{1/2}$,  and rearranging the terms, the equation in the aforementioned approximation reads

\beq
\frac{d^2\varphi}{d\tau^2}+\frac{\omega_j}{R I_c} \frac{\hbar}{2e}  \frac{d\varphi}{d\tau}+ \sin (\varphi)=\frac{I_B}{I_c} + 
\frac{S_0}{I_c} \sin(\frac{\Omega}{\omega_j} \tau +\varphi_0)+ \frac{S_N}{I_c}\tilde{\xi}(\tau).
\label{eq:josnormC}
\eeq

In Eq.(\ref{eq:josnormC}) we have defined $S_N=\sqrt{\omega_j D}$ the intensity of noise current, while the correlator in these normalized units reads 
\beq
< \tilde{\xi}(\tau) \tilde{\xi}(\tau') > = 2 \delta(\tau-\tau')
\eeq 

Defining $\gamma=I_B/I_c$ the normalized bias current, $\alpha = (\omega_j/RI_c) (\hbar/2e)$ the normalized dissipation, $\varepsilon=S_0/I_c$ the normalized signal amplitude and $\varepsilon_N=(S_N/I_c)^2$ the normalized noise intensity, Eq.(\ref{eq:josnormC}) reads:

\beq
\frac{d^2\varphi}{d\tau^2}+\alpha  \frac{d\varphi}{d\tau}+ \sin(\varphi)=\gamma + \varepsilon \sin(\omega \tau +\varphi_0)+ \sqrt{\varepsilon_N} \tilde{\xi}(\tau).
\label{eq:josnorm}
\eeq

A washboard potential is associated to Eq.(\ref{eq:josnorm}) that, for $\gamma < 1$, gives rise to a barrier \cite{Benjacob}

\beq
\Delta U(\gamma)=2  [\sqrt{1-\gamma^2} - \gamma \cos^{-1}(\gamma)].
\label{eq:barrier}
\eeq

\noindent The schematic of the physics of the device is depicted in Fig. \ref{fig:potential} as a jump over 
an activation barrier. When the system overcomes the energy barrier $\Delta U$, it switches from 
the locked state to a running state that is associated with a finite voltage \cite{risken}, 
hence it is possible to measure the escape time \cite{taumeasure}. For overdamped JJ the voltage step is very smooth,
and it is difficult to define the escape from the local solution. Therefore the detection efficiency of overdamped JJ drops down.

\begin{figure}
\centerline{\includegraphics[scale=0.5]{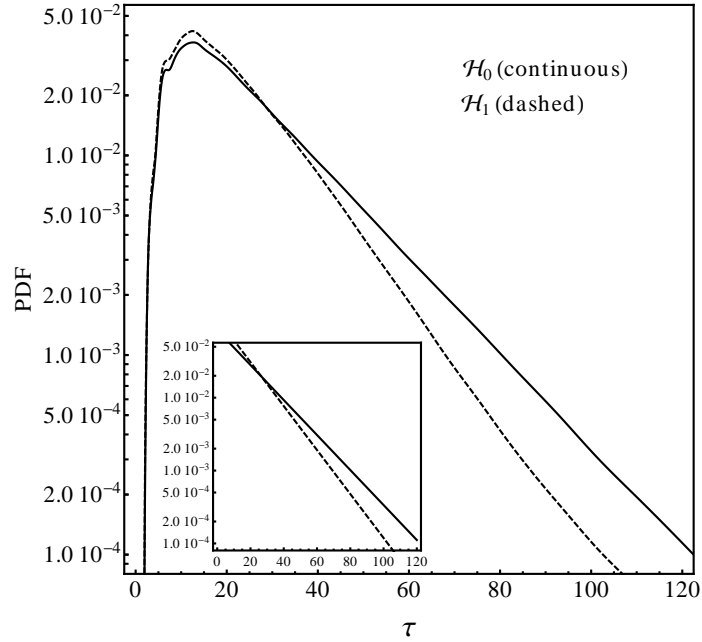}}
\caption{Distribution of the escape times with the sinusoidal signal (dashed line) and 
without (continuous line). The main effect of the signal for this range of the parameters is to change the slope of the distribution.
The inset shows the asymptotic fit with exponential distributions.
The resulting decay rate in presence of (without) the signal is $0.069$ (resp. $0.055$).
Parameters of the simulations are: $\gamma = 0.5$, $\alpha = 0.05$, $\varepsilon_N = 0.07$. Moreover, when the signal is present,  $\varepsilon = 0.1$, $\varphi_0 = 0$ and $\omega = 0.035$.}
\label{fig:esctimedistr1}
\end{figure}

\begin{figure}
\centerline{\includegraphics[scale=0.5]{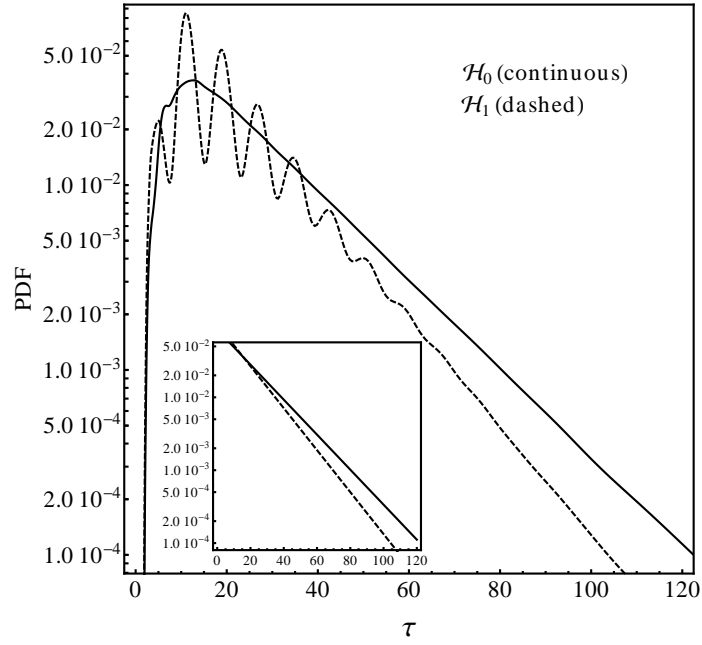}}
\caption{Distributions of the escape times with the sinusoidal signal (dashed line) and without (continuous line).
The main effect of the signal for this range of the parameters is to introduce oscillations on the escape time distribution.
The inset shows the asymptotic fit with exponential distributions. 
The resulting decay rate in presence of (without) the signal is $0.066$ (resp. $0.055$).
Parameters of the simulations are: $\gamma = 0.5$, $\alpha = 0.05$, $\varepsilon_N = 0.07$. Moreover, when the signal is present,  $\varepsilon = 0.1$, $\varphi_0 = 0$ and $\omega = 0.8$.}
\label{fig:esctimedistr2}
\end{figure}

The main idea of signal detection is to collect the escape times to discriminate between two situations: 
\emph{i}) the exit is caused by the presence of pure noise (no signal is present, $S_0=0$); 
\emph{ii}) the exit is caused by the joint action of noise and a sinusoidal excitation (the signal is present, $S_0\neq0$).
Indeed escape time distributions are highly sensitive to the signal amplitude, as shown since the pioneering experiments \cite{Devoret}. 
Typical escape time Probability Density Functions (henceforth PDFs) are shown in Figs. \ref{fig:esctimedistr1}, \ref{fig:esctimedistr2} for two different sets of parameters.
The random sequence of escape time  is obtained by numerical integration of stochastic differential equation (\ref{eq:josnorm}), accumulating the first-passage-time to cross the maximum
of  the potential barrier (\ref{eq:barrier}) and taking into account the corrections of Ref. \cite{Mannella01}. 

\begin{figure}
\centerline{\includegraphics[scale=0.5]{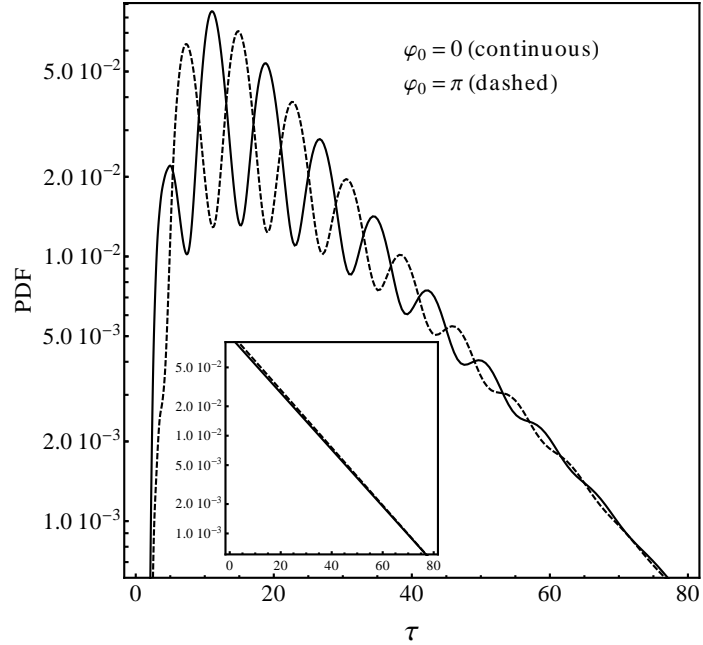}}
\caption{Distributions of the escape times with a signal with two initial phases, $\varphi_0=0$ and $\varphi_0=\pi$. 
The inset shows the asymptotic fit with exponential distributions, it is evident that the exponential decay rate is almost the same.
Indeed the resulting decay rate for $\phi_0=0$  ($\phi_0=\pi$)  is $0.067$ ($0.068$).
The other parameters of the simulations are: $\gamma = 0.5$, $\alpha = 0.05$, $\varepsilon_N = 0.07$, $\varepsilon = 0.1$ and $\omega = 0.8$.}
\label{fig:esctimedistr3}
\end{figure}

A solid line denotes the escape time without signal ($S_0=0$) and the dashed line denotes the escape 
aided by a sinusoidal forcing ($S_0\neq 0$). 
A possible signal detection strategy based on the sample mean of the escape times \cite{filatrella10} only measures the average escape time, that is essentially the slope shown in the inset of  Figs. \ref{fig:esctimedistr1}, \ref{fig:esctimedistr2}. 
On the other hand, Fig. \ref{fig:esctimedistr2} makes it clear that the two distributions (with and without the sinusoidal signal) are, for some parameters values, very different in shape, not just in the mean value. 
This difference in shape is exploited in Section III, where it is shown that a more refined analysis can lead to better performances for signal detection. 
Moreover, in Fig. \ref{fig:esctimedistr3} it is shown the effect of the signal (initial) phase. 
The drastic change in the distribution form, while the slope remains almost constant, 
again demonstrates the need of a more refined analysis, to achieve signal phase estimation by using the escape times, as discussed in Section IV C.

The escape time distribution not only depends upon the signal, but also on the physical JJ features. 
The JJ parameters that can be set in the experiments to achieve best detection are the electrical characteristics of the physical JJ 
($C$, $R$, $I_c$) and the external bias current $I_B$. 
Moreover, if the signal is recorded, also the speed at which the recorded signal $S(t)$ is physically applied is a tunable parameter. 
A blind search among such multidimensional parameter space would be numerically prohibitive and of little physical significance; 
we therefore discuss in Section IV the physical meaning of the parameters to restrict the plausible region where to seek for best performances.

\section{Strategies for JJ threshold detection}

The JJ activation energy barrier defined in Eq. (\ref{eq:barrier}) refers to an unperturbed junction, i.e. to the case $S_0=0$ in Eq. (\ref{eq:josnorm}). 
The addition of a deterministic sinusoidal signal of amplitude $S_0$ results in an oscillation of the barrier that affects the average escape time. 
The deviations of such average escape time shown in Figs. \ref{fig:esctimedistr1}, \ref{fig:esctimedistr2} can be used to 
infer the presence of a signal. Inspection of Figs. \ref{fig:esctimedistr1} and \ref{fig:esctimedistr2} discloses an evident exponential 
decay of the escape time distribution. 
Such behavior has been previously exploited (see \cite{filatrella10}) with a straightforward SM detection strategy based on the mean escape time.
Indeed it can be shown that if the escape times are exponentially distributed in both cases, $S_0=0$ and $S_0 \neq 0$, then the SM detection strategy is an optimal strategy \cite{Shao}.
In the same paper \cite{filatrella10} the Kumar-Carrol index $d_{KC}$  \cite{Kumar} for different SNR (we recall that the SNR is related to the ratio 
$S_0/S_N=S_0/\sqrt{\omega_j D}$) has been used as a simple (heuristic) indicator of the detector performances.
We remind that the index $d_{KC}$ depends upon the operating frequency of the applied signal and the extrema are influenced by the noise intensity \cite{yu,filatrella10}, a peculiarity of stochastic resonance \cite{Gammaitoni89,mcnamara89}.

When the PDFs are not exponential (and not explicitly known, as it is the case for underdamped JJ) we use an accurate and fairly common technique to determine a detection strategy and the related detector, both based on likelihood maximization. Such a decision criterion, based on LRT, is optimal \cite{Shao} in a sense that is clarified below. Detectors based on LRT employ the knowledge of the full probability 
distribution of the random escape with and without the signal; therefore the shape of PDFs can be properly handled to improve the performances. 

Unfortunately, the escape time distributions are not theoretically known  for the system described by Eq. (\ref{eq:josnorm}). 
Even in the case $S_0=0$ the Arrhenius law is approximately valid for rare escapes \cite{devoret85} (in the unperturbed oscillator timescale $\omega_j$), 
while for fast escapes \cite{silvestrini88}
(that are interesting for signal analysis) only approximated analytic estimates exist. 
When the signal is applied the 
knowledge of the escape time distributions is even poorer, and essentially limited to the overdamped case \cite{guentz}. 
Our solution is based on a semi-analytic approach, so that the analytic results are used as a guess for a numerical procedure. 
In the following we describe the proposed LRT embedded in the general framework
of statistical decision theory.

\subsection{Statistical Decision Theory and LRT Detector}

To properly define the detection strategies,
it is usual to formalize the problem as a binary hypothesis test:
\[
\begin{array}[c] {cl}
\mathcal{H}_0: & \text{sinusoidal signal is absent}\\
\mathcal{H}_1: & \text{sinusoidal signal is present}
\end{array}
\]
For this decision problem two different error probabilities arise:
\begin{itemize}
\item the \emph{false alarm probability} $P_f$, also called Type I error probability, i.e. the probability to decide for the hypothesis $\mathcal{H}_1$ when $\mathcal{H}_0$ is true;
\item the \emph{miss probability} $P_m$, also called Type II error probability, i.e. the probability to decide for the hypothesis $\mathcal{H}_0$ when $\mathcal{H}_1$ is true.
\end{itemize}
We start by considering the case in which the JJ normalized parameters ($\alpha$ and $\gamma$) and the normalized noise standard deviation $\varepsilon_N$ are perfectly known and do not depend on the particular hypothesis in force. We also assume that the signal parameters (i.e. $\varepsilon$, $\omega$ and $\varphi_0$) are known under $\mathcal{H}_1$ hypothesis.
In this setup the Neyman-Pearson lemma \cite{helstrom} identifies the LRT as the optimal detection strategy, for it minimizes, among all possible tests, the miss probability $P_m$ at a fixed false alarm level $P_f$.
Thus if we collect $N$ escape times $\underline{\tau} = \left\{\tau_i\right\}_{i\in[1, N]}$, supposed to be independent and identically distributed, the test statistic can be written as:
\beq
\displaystyle{\prod_{i = 1}^{N}\frac{f_1(\tau_i)}{f_0(\tau_i)}} 
\begin{array}[c]{c}
{\mathcal{H}_1}\\{>} \\
{<}\\{\mathcal{H}_0} 
\end{array}
\zeta^{\prime},
\label{eq:likeT}
\eeq
where $f_{0,1}(\cdot)$ are the PDFs of the escape times under the hypothesis $\mathcal{H}_{0,1}$, while $\zeta^{\prime}$ is a suitable threshold selected to return a fixed false alarm level.
To simplify the computation of the statistic (\ref{eq:likeT}), it is possible to compare the normalized natural logarithm of the likelihood ratio with a threshold $\zeta = \log(\zeta^{\prime})/N$:
\beq
\Lambda(\underline{\tau}) = \frac{1}{N}\displaystyle{\sum_{i = 1}^{N}\log \left[ \frac{f_1(\tau_i)}{f_0(\tau_i)} \right]}
\begin{array}[c]{c}
{\mathcal{H}_1}\\{>} \\
{<}\\{\mathcal{H}_0} 
\end{array}
\zeta.
\label{eq:loglikeT}
\eeq
The advantage of Eq. (\ref{eq:loglikeT}) is that the statistic $\Lambda(\underline{\tau})$ can be computed as the sample mean of the random samples 
$\underline{\mathcal{L}} = \left\{\mathcal{L}_i\right\}_{i\in[1, N]}$, that are obtained from the escape times via the \emph{optimal} (in the Neyman-Pearson sense), non-linearity
\beq
\mathcal{L}_i = \log \left[ \frac{f_1(\tau_i)}{f_0(\tau_i)} \right].
\label{eq:locloglik}
\eeq
Equation (\ref{eq:locloglik}) contains the information of both PDFs  $f_1(\cdot)$ and $f_0(\cdot)$.
Unfortunately, for underdamped JJ, an exact closed form of these PDFs is still unknown. 
We have seen in Figs. \ref{fig:esctimedistr1} and \ref{fig:esctimedistr2} that there are regimes in which both the densities follow, with a good approximation, an exponential law. 
As anticipated, in this case the SM detector is nearly optimal. Indeed it is straightforward to see that for exponential distributions the decision statistic in
 Eq. (\ref{eq:loglikeT}) becomes:
\beq
\mathcal{A}(\underline{\tau}) = \frac{1}{N}\displaystyle{\sum_{i = 1}^{N}\tau_i}
\begin{array}[c]{c}
{\mathcal{H}_1}\\{>} \\
{<}\\{\mathcal{H}_0} 
\end{array}
\zeta.
\label{eq:smT}
\eeq
We note that if the average escape time under hypothesis $\mathcal{H}_0$ is larger than the same quantity under $\mathcal{H}_1$,  
the sign in the Eq. (\ref{eq:smT}) should be reversed.

In other regimes we estimate both PDFs $f_1(\cdot)$ and $f_0(\cdot)$ using a non-parametric statistical technique such as the \emph{Kernel Density Estimation} (KDE)
\cite{silverman}. 
Thus, by means of a large number of samples $\tau_i$ ($5\cdot10^5$ trials), 
obtained via a Monte Carlo simulation of the escape process of Eq. (\ref{eq:josnorm}), in both cases $S_0=0$
and $S_0 \neq 0$ we retrieve a tight estimate $\widehat{f}_j(\cdot)$ of the unknown PDFs $f_j(\cdot)$. Further details about the KDE are given in the Appendix A.

\begin{figure}
\centerline{\includegraphics[scale=0.5]{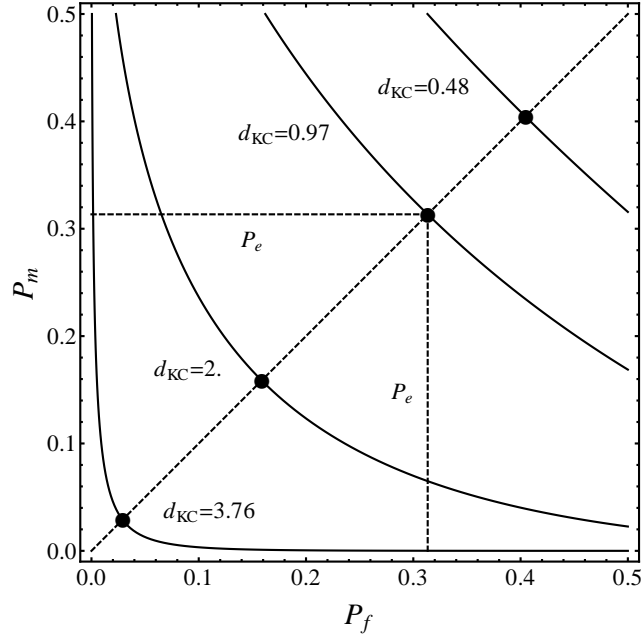}}
\caption{Typical ROCs of a JJ based detector under the hypothesis of complete parameter knowledge for different values of $d_{KC}$ 
(i.e. of $\varepsilon$).  
Other relevant simulation parameters are: $\gamma = 0.3$, $\alpha = 0.05$, $\varepsilon_N = 0.07$. 
Moreover, when the signal is present, $\varphi_0 = 0$ and $\omega = 0.8$.}
\label{fig:ExamplePe}
\end{figure}

We are now in a position to  compute the Receiver Operator Characteristic (ROC)  of the test statistic, that is the plot of $P_f$ vs. $P_m$ for different values of $\zeta$ \cite{notarock}. 
A ROC example is presented in Fig. \ref{fig:ExamplePe}, 
in which it is evident the trade-off between the two error probabilities.
To simplify the performance analysis of the detector, we consider the intersection 
between the ROC and the bisector of the first quadrant angle, that is very close to the point of ROC curve with the
 minimum distance to the axis origin. 
In this point  $P_f = P_m$, and we can unambiguously define the error probability $P_e$ that is representative of the detector behavior. 
The main advantage of this formulation resides in its simplicity and, in many cases, it is also a good approximation of the minimum Bayesian 
error probability when the prior probabilities of the two hypothesis $\mathcal{H}_0$ and $\mathcal{H}_1$ are considered equal.
The error probability $P_e$ gives a rigorous assessment of the detector's performance, and 
can be related to the heuristic Kumar-Caroll index $d_{KC}$ by an inequality, as elucidated in Appendix B. 

\subsection{Escape Time Acquisition Strategies}

The signal to noise mixture can be applied to the JJ in different ways \cite{filatrella10} to acquire the escape time sequence $\underline{\tau}=\{\tau_i\}_{i=1}^N$.
Indeed, when the JJ  switches to the running state (see Fig. \ref{fig:potential}) 
it is necessary to shield the junction from the signal to reset the static state and to apply again the signal. There is a variety of methods to reset the system. 
If the frequency of the signal is perfectly known it is in principle possible to reapply the signal always with the same initial phase. This acquisition strategy is called coherent detection. To apply again the signal with the same initial phase some fraction of the signal is lost waiting for the correct time to restart the process.
In fact the correct initial times read: 

$$
t^{(r)}_i=\frac{2\pi}{\omega} \lceil \frac{\omega \tau_i}{2\pi} \rceil + t^{(r)}_{i-1}.
$$

\noindent With this acquisition method the escape PDF shows a striking dependence from $\varphi_0$, as elucidated in Fig. \ref{fig:esctimedistr3}. 

Another possibility, if the frequency is unknown, is to reapply the signal with any phase it might have after the reset procedure,
and therefore with an essentially 
random initial value of $\varphi_0$ (incoherent strategy). 
The PDF of the escape time obtained in this case, essentially the average over $\varphi_0$ of the escape densities computed
in the coherent strategy case, looses much of the information carried by the initial phase.

\subsection{Filter Bank strategy and Phase Estimation}

In the detection theory briefly summarized  above, the involved parameters are supposed to be known: the LRT method can only be employed if the signal 
is supposed to have a known phase.  
Unfortunately in real scenarios this condition is rarely fulfilled and the PDF of the escape times should be considered a function of an unknown parameter vector $\underline{\theta}$.
It is necessary to properly manage the lack of information about $\underline{\theta}$ to contrive a detection rule that 
minimizes the deterioration of the performances for unknown parameters. In this subsection we show how one can retrieve with a filter bank some information about the phase of the signal. The main problem arises because the signal itself might be present or not, and therefore one cannot employ a simple maximum likelihood procedure to determine the best guess for the parameter value.
Two  popular approaches to the detection with unknown parameter are currently used in signal processing:
the {\em Averaged} LRT (ALRT) and the {\em Generalized} LRT (GLRT).
The ALRT, based on the Bayesian Theory, consists in averaging the likelihood ratio on the unknown parameters over the density functions corresponding to all the values of the unknown parameter; the resulting best guess is then employed in a Neyman-Pearson criterion to decide the signal presence.
GLRT estimates the unknown parameters via a maximum likelihood approach to select a single most likely phase. Such phase is substituted in the PDF expression of the LRT to decide about the signal presence. 

\begin{figure}
\centerline{\includegraphics[scale=.7]{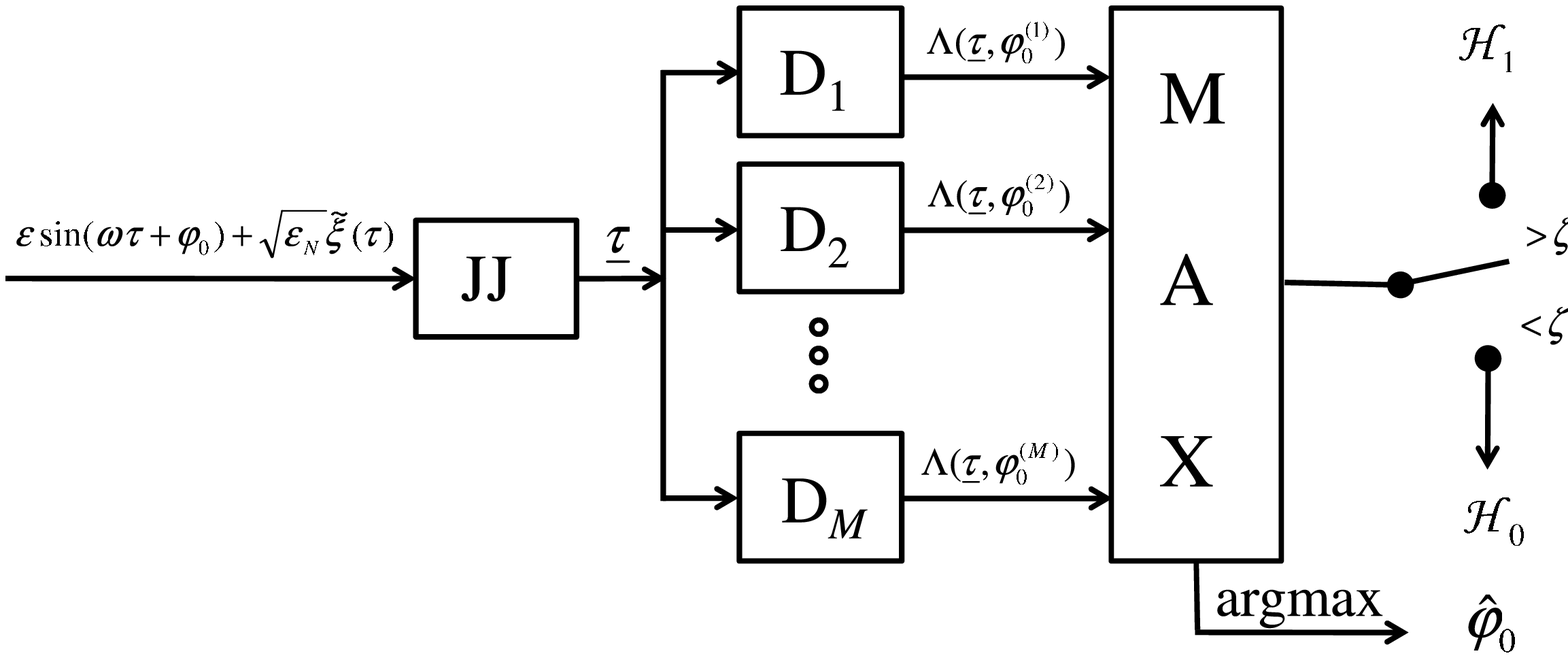}}
\caption{Schematic of the filter bank.The JJ block indicates the physical device giving the
escape time, while the $D_i$ blocks represent the filter bank in which the elements are tuned on different values of the unknown parameter (in this case the initial phase $\varphi_0$). The maximization block (MAX) output is compared with a suitable 
threshold $\zeta$ to perform the decision test while $\argmax$ gives the maximum likelihood phase estimation.}
\label{Fig:BancoFase}
\end{figure}

In  this paper, as a paradigmatic example, we suppose that the unknown parameter is the initial phase, i.e.  $\underline{\theta} = \varphi_0$.
Without any {\em a priori} information, the ALRT approach is substantially equivalent to apply the incoherent acquisition strategy.
Unfortunately (as is shown below) this approach looses too much information about the signal presence. Thus we focus on the GLRT, that can be implemented via a Filter Bank (see Fig. \ref{Fig:BancoFase} for a pictorial scheme) . 
The filter bank jointly performs the estimation and the detection of the unknown parameter  $\varphi_0$.
In a preliminary step we should determine a sampling of the relevant parameter space $\{\varphi_0^{(i)}\}_{i=1}^M$ such that it is suitably covered the interval $[-\pi,\pi]$.
Each detector of the bank $LRT_i$, tuned on the value $\varphi_0^{(i)}$ and designed as described in subsection III A, 
elaborates the sequence of escape times $\underline{\tau}$ that are acquired with a coherent strategy to obtain a vector of likelihood ratios $\{\Lambda(\underline{\tau},\varphi_0^{(i)})\}_{i=1}^M$. 
We recall that the phase $\varphi_0$, as depicted in Fig. \ref{fig:esctimedistr3}, influences the PDF of $\underline{\tau}$ \emph{only} when the signal is present. 
Thus the maximization of the Likelihood under $\mathcal{H}_1$ coincides with maximizing $\Lambda(\underline{\tau},\varphi_0)$ and the test (\ref{eq:loglikeT}) becomes 
\beq
\max_{i=1,\dots,M}\left[ \Lambda(\underline{\tau},\varphi_0^{(i)})) \right]
\begin{array}[c]{c}
{\mathcal{H}_1}\\{>} \\
{<}\\{\mathcal{H}_0} 
\end{array}
\zeta.
\label{eq:GloglikeT}
\eeq
The value of $\varphi_0^{(i)}$ that maximizes $\Lambda(\cdot)$ constitutes an estimate of the initial phase
\beq
\widehat{\varphi}_0 = \argmax_{i=1,\dots,M}\left[ \Lambda(\underline{\tau},\varphi_0^{(i)})) \right].
\label{eq:GloglikeEst}
\eeq
To optimize the performance of such strategy, one should design the filter bank with the appropriate choice of 
both the number $M$ of the detector and the values $\varphi_0^{(i)}$. 
We propose to place the values $\varphi_0^{(i)}$ uniformly spaced in $[-\pi,\pi]$ by sake of simplicity. Also, we have considered that the JJ
 behavior for a particular phase value is not too different from another one \cite{filatrella10}.
The number $M$ should be determined recognizing that few detectors offer a poor precision in phase estimation and in detection effectiveness, 
while too many detectors significantly increase the false alarm probability. A more precise analysis of the bank to quantitatively balance the two sides is out of the scope of this paper.

\section{Simulation and Numerical Results}

In this Section we present extensive Monte Carlo simulations that show the behavior of a detector based on a JJ. 
As evident in Eq. (\ref{eq:loglikeT}), LRT employs a fixed number $N$ of escape times and it is \emph{optimal} (in the Neyman-Pearson sense), under this condition. On the other hand the time interval $T_{obs}$ to collect $N$ escape times is random and strongly depends on both the signal properties and the JJ parameters.
Indeed, when the signal is absent (hypothesis $\mathcal{H}_0$), the mean value of the escape times 
$\mu_0 = E[\tau|\mathcal{H}_0]$ , that is a function of $\gamma$, $\alpha$ and $\varepsilon_N$,  reads
\beq
E[T_{obs}|\mathcal{H}_0] = N \mu_0.
\label{eq:IncoTime}
\eeq
We observe that also in presence of small signals the escape rate remains roughly unchanged $\mu_1  = E[\tau|\mathcal{H}_1] \simeq \mu_0 $. 

To perform a fair comparison
under different operative conditions, we fix the mean time interval under the $\mathcal{H}_0$ hypothesis and use the corresponding number of escape times $N$. 
This strategy guarantees a mean duration time of the acquisition stage in any practical situation ($\mathcal{H}_0$ and $\mathcal{H}_1$).
Moreover, while the mean time of $T_{obs}$ increases linearly with $N$, its standard deviation is proportional to $\sqrt{N}$, so the dispersion around the mean value becomes less significant for increasing values of $N$.

Equation (\ref{eq:IncoTime}) is strictly true only for the incoherent acquisition. 
Indeed in the coherent case we have to add to $\mu_0$ the mean time needed to assure that the incoming signal is applied to JJ 
with the same initial phase. So the approximate relation
\beq
\displaystyle{E[T_{obs}|\mathcal{H}_0] \approx \left\{
\begin{array}{lc}
N \left(\frac{2 \pi}{\omega}\right), & \mu_0 \leq \frac{\pi}{\omega}\\
N \left(\mu_0 + \frac{\pi}{\omega}\right), & \mu_0 > \frac{\pi}{\omega}
\end{array}\right.}
\label{eq:CoheTime}
\eeq
reveals that the normalized frequency $\omega$ also influences the observation time. We have proven by numerical simulations (not shown here) that this effect 
is negligible (in the interesting range of parameters), so we use the simple approximate Eq. (\ref{eq:CoheTime}) in the remaining part of the paper.

\subsection{LRT Performance Improvements and Signal Frequency Analysis}

In this Subsection we investigate the dependence of the error probability $P_e$ as a 
function of the normalized signal frequency $\omega$,while in the next Subsection we carefully analyze the dependence of the performances upon the junction parameters. The distinction is somewhat artificial, inasmuch the normalized frequency $\omega$ also depends upon the junction parameters:  $\omega = \Omega/\omega_j=\Omega/[2eI_c/(\hbar C)]^{1/2}$. 
However, in view of some emerging physical and signal detection properties,  
we prefer to focus the analysis of the signal frequency and phase in this Subsection, and postpone the analysis of the other junction parameters.

\begin{figure}
\centerline{\includegraphics[scale=0.5]{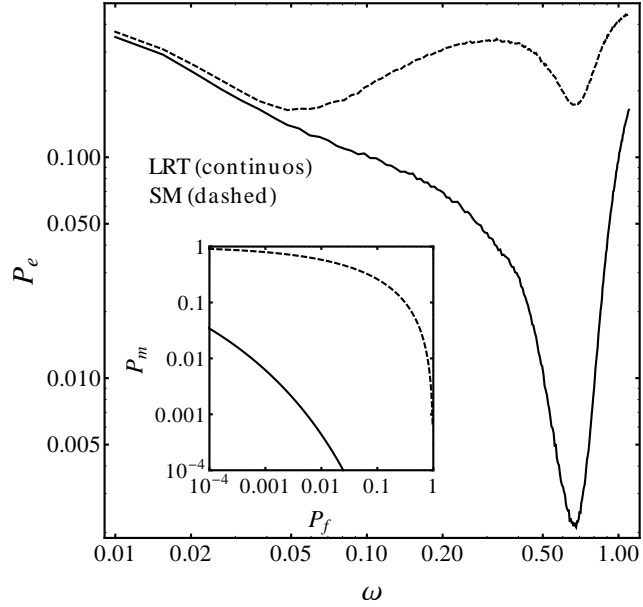}}
\caption{Detection probability as a function of the applied signal frequency. Parameters of the system are $\gamma=0.8$, $\alpha = 0.05$, $\varepsilon_N = 0.0175$, $\varepsilon=0.05$ and $\varphi_0 = 0$. The simulations are performed  setting the mean observation time under $\mathcal{H}_0$,  $E[T_{obs}|\mathcal{H}_0] =2000$. 
In the inset the ROCs computed for both the detectors at a frequency close to geometric resonance $\omega_{res}\approx \omega = 0.7$.}
\label{fig:det_vs_omega}
\end{figure}

\begin{figure}
\centerline{\includegraphics[scale=0.5]{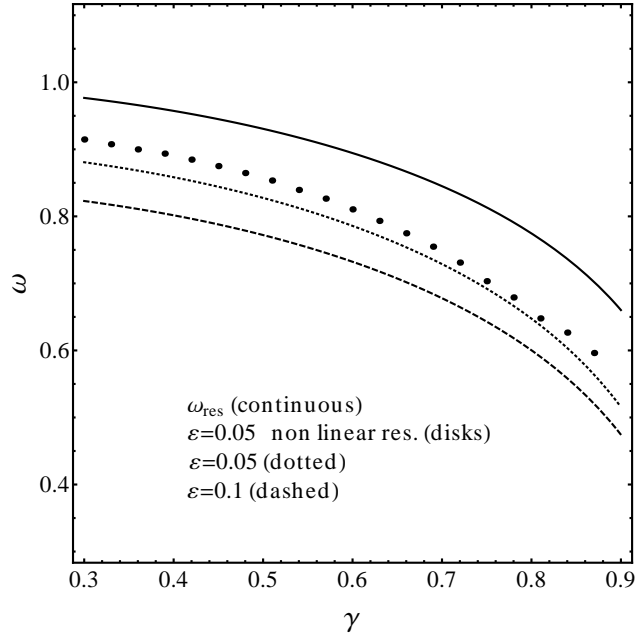}}
\caption{Plot of frequency $\omega$ where a minimum of $P_e$ occurs as a function of the bias current $\gamma$ for LRT detection strategy.
Dashed line refer to a signal with $\varepsilon=0.1$ and a noise with $\varepsilon_N=0.07$, while dotted line to $\varepsilon=0.05$ and a noise with $\varepsilon_N=0.0175$ 
(see also Fig. \ref{fig:omega_gamma}). For sake of comparison we also show (continuous line) 
the resonant frequency of Eq. (\ref{eq:omegares})  corresponding to vanishingly small $\varepsilon$ and the non linear deterministic resonance curve (dots) in the case $\varepsilon=0.05$. 
Other relevant parameters of the system are: $\alpha = 0.05$ and $\varphi_0 = 0$. The simulations are performed setting the mean observation time under $\mathcal{H}_0$,  $E[T_{obs}|\mathcal{H}_0] =2000$.}
\label{fig:optres}
\end{figure}

The dependence of the detection properties upon the signal frequency is shown in Fig. \ref{fig:det_vs_omega}, where we compare the SM of escape times (broken lines) with the more refined LRT (solid line). We also show in the inset the ROC of both the tests at a single frequency.
In spite of the approximations used to obtain Eq. (\ref{eq:locloglik}) the improvements of the LRT detection performances are significant.
Along the bisector of Fig. \ref{fig:det_vs_omega} inset the probability of dismissal $P_m$ decreases of a factor $\simeq 10^2$ 
from the simple average to the LRT. Since the two curves are computed for the same signal duration, one can also deduce that the
LRT allows for a decrease of $\simeq 10^4$ of the signal length keeping the same false alarm level. 
It is also evident from Fig. \ref{fig:det_vs_omega} that for all driving frequencies the LRT overperforms the SM detector.
Both strategies share a pronounced dip at the geometric resonance.
Such  frequency is not exactly $\omega_j$, for the tilted washboard potential of Eq. (\ref{eq:barrier})
exhibits a dependence of the resonant frequency  upon the bias current of the type \cite{BarPat}:
\beq
\omega_{res} \simeq \left(1-\gamma^2\right)^{1/4}
\label{eq:omegares}
\eeq

\noindent that is mirrored in the performances of the detector. 
The resonant condition Eq. (\ref{eq:omegares}) is obtained through linearization of Eq. (\ref{eq:josnorm}) for small signal amplitude $\varepsilon=S_0/I_c$.
For finite signal amplitude the oscillations explore the nonlinear part of the curvature that is not captured by the second order Taylor expansion behind 
Eq. (\ref{eq:omegares}). The correction, or rectification effect, accounts for the deviation from Eq. (\ref{eq:omegares}) of the resonant 
frequency at finite signal amplitude, $\omega_{res} = \omega_{res}(\varepsilon)$.
In  Fig. \ref{fig:optres} is shown the non linear JJ plasma resonance curve  \cite{nonlinear};
the higher order deterministic correction of the plasma frequency approaches the optimal  
working frequency for the detection. The displayed non linear  plasma frequency
is the stable branch in the  regime where the resonance curve becomes a multivalued function of the frequency \cite{Landau}.
As a consequence we speculate that best detection performances are attained in the strongly non linear distortion regime. 
In our opinion this result mirrors the similar findings of Ref. \cite{superamp} concerning the location of the best amplification region in  the parameter space.
It is noteworthy that such frequency optimization refinement is needed to ensure that the detector's resonance matches 
the signal frequency. We conclude the analysis of the geometric resonance affirming that both SM and LRT detectors 
exhibit the dependence predicted by Eq. (\ref{eq:omegares});
there exists a suitable neighborhood of $\omega_{res}$ that is one of the  best region for detection purposes, with a small correction for the finite signal amplitude.

The SM strategy reveals that a second interesting region occurs at a lower normalized signal frequency, where another resonance appears.
The position of such a resonance dip depends upon the phase and the temperature  \cite{filatrella10}, 
see Figs. \ref{fig:det_vs_omega},\ref{fig:cont_omega_gamma}(a), \ref{fig:omega_gamma}(a).
In fact in Ref.\cite{filatrella10} it has been found that a region of optimal detection is pinpointed if the potential 
well barrier (tuned by $\gamma$), the normalized signal frequency $\omega$ (the signal frequency $\Omega$ 
divided by the Josephson frequency $\omega_j$) and the noise  intensity $D$ are connected by the relation

\beq
\omega_{SR}=\frac{\Omega_{SR}}{\omega_j}=\frac{\tau_0}{2\pi C(\varphi_0)}
 \exp\left\{\frac{2 I_c^2} {\omega_j D}\left [\sqrt{1-\left (\frac{I_B}{I_c}\right)^2} - \frac{I_B}{I_c} 
\cos^{-1}\left(\frac{I_B}{I_c}\right)\right] \right\}.
\label{eq:tau_barrier}
\eeq
Below the stochastic resonance frequency the PDFs (in both hypotheses) are very similar to those reported in Fig. \ref{fig:esctimedistr1}, while above
such frequency the PDF (under ${\cal H}_1$) develops oscillations similar to those reported in Fig. \ref{fig:esctimedistr2}. This explains the disappearance of stochastic resonance (\ref{eq:tau_barrier}) in the LRT detection framework, for LRT exploits the PDFs oscillations and does not deteriorate above the frequency (\ref{eq:tau_barrier}). 
Thus the paradoxical increase of the performances at higher noise level for the SM is solved by the observation that the improvement obtained at the $\omega_{SR}$ frequency 
is outperformed by the choice of a more refined LRT detection strategy that takes into account the PDFs oscillations.  In fact, see Figs. \ref{fig:det_vs_omega}, the SM detector performances are always worse than the LRT ones, confirming the general idea that stochastic resonance is a consequence of a suboptimal detection scheme \cite{galdi}. 
The practical consequence is that synergetic effects leading to stochastic resonance between noise and signal in nonlinear devices can only be exploited in suboptimal strategies, while in optimal detection strategies noise should only be reduced as much as the experimental set up allows.

\subsection{Physical Considerations on JJ Parameter Optimization}

In this Subsection we study the detector performances as a function of JJ parameters. 
Furthermore we transform the results obtained with simulations of the normalized 
Eq. (\ref{eq:josnorm}), and analyzed with the methods of Sec. III, into prescriptions  for the physical parameters of an actual JJ. 
The system depends upon four normalized parameters that can be tuned, with an appropriate choice of the JJ physical parameters,
 to obtain the best performances. The JJ features that can be modified are:

\begin{enumerate}
\item The Josephson frequency $\omega_j=[2eI_c/(\hbar C)]^{1/2}$ (and hence the capacitance $C$ and the critical current $I_c$) selects the time scale of the device and must be chosen to maximize the performances of the device. One should compromise between two different requirements: fast to speed up the detection, but still slow enough to allow the available electronic to properly work. The critical current is typically constraint in the range $1 \mu A \le I_c \le 10 mA$, while the capacitance is in the range $1-1000 pF$. The two quantities are not fully independent, for both depend upon the distance between the superconducting electrodes and the junction area. The critical current can also be decreased by an external magnetic field.
Typically, the available range of $\omega_j$ is about $10-1000 GHz$.

\item  The applied physical current $I_B$ can be assumed positive (for the symmetry of the problem a negative bias value would just revert the phenomena) and below the critical current to have two solutions (see Eq. (\ref{eq:barrier})), $0 \le I_B \le I_c$. In normalized units the interval reads $0 \le \gamma \le 1$. 

\item The applied signal amplitude $S_0$ cannot be amplified without introducing a further noise component. It is therefore convenient to set instead the critical current to have the more appropriate $\varepsilon = S_0/I_c$ normalized signal. It is worth noticing that the normalized intensity of the noise 
$\varepsilon_N = \omega_j D/I_c^2$ also depends upon the critical current $I_c$, and can therefore be also tuned, while obviously the SNR $\propto \varepsilon/\sqrt{\varepsilon_N} = S_0/\sqrt{\omega_j D}$ can not. As already mentioned the critical current of a Josephson junction is typically constraint in the range $1 \mu A \le 10 mA$, and therefore it is not possible to freely choose the normalized signal amplitude. Also, it is important to notice that if the critical current $I_c$ is used to tune the value of the normalized signal, the system frequency $\omega_j$ can still be modified via the capacitance $C$. 

\item The resistance $R$ sets dissipation through the normal electron channel, parallel to the tunnel Josephson element. Dissipation enters in Eq. (\ref{eq:josnorm}) through the normalized parameter $\alpha= (\omega_j/RI_c) (\hbar/2e)$. The resistance is constrained, inasmuch the product $I_c R$ depends upon the material: $I_c R = \Delta/2e$. The energy gap $\Delta$ for type I traditional superconductors is in the range of $meV$. 
The normalized parameter $\alpha$ is limited in the interval $0 < \alpha < 1$. The lower value is due to physical reasons (all quantities are positive), while the upper value is necessary to have a so-called hysteretic junction with two states (detection depends upon the possibility to detect the switch between these two states). 
Actual values for viable JJ are narrowed in the range $ 0.001 \le \alpha \le 0.1$, while the range $ 0.1 \le \alpha \le 1 $ corresponds
to the moderately damped regime where retrapping occurs, making it difficult to detect the escape from  the static solution \cite{Fenton08}.
Finally, dissipation can be increased shunting the junction with an external resistor.
\end{enumerate}

One can interpret the physical parameters $R$, $C$, $I_c$ and $I_B$, and the corresponding normalized parameters $\alpha$, $\omega$, $\varepsilon$, and $\gamma$ in the following way. The normalized frequency $\omega=\Omega/\omega_j$ can be chosen to drive the JJ with the most appropriate (for signal detection) 
frequency, and this sets the ratio of the critical current $I_c$  and 
capacitance $C$. The bias current $\gamma$ that can be varied through the external bias current $I_B$ sets the potential well of the system, see Eq. (\ref{eq:barrier}), while the resistance $R$ can be adjusted to have the most appropriate value of dissipation.

\begin{figure}
\centerline{(a)}
\centerline{\includegraphics[scale=0.5]{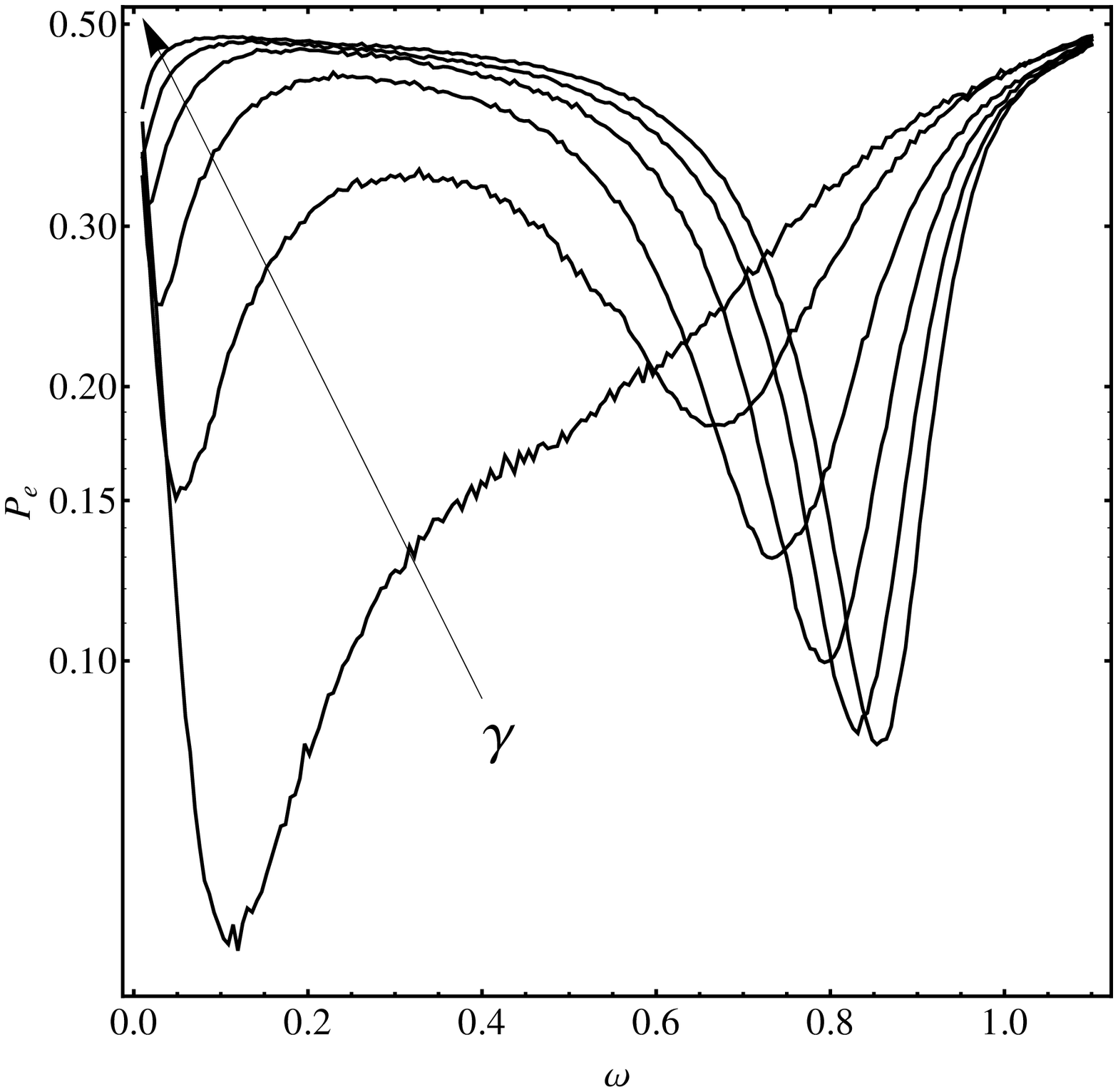}}
\centerline{(b)}
\centerline{\includegraphics[scale=0.5]{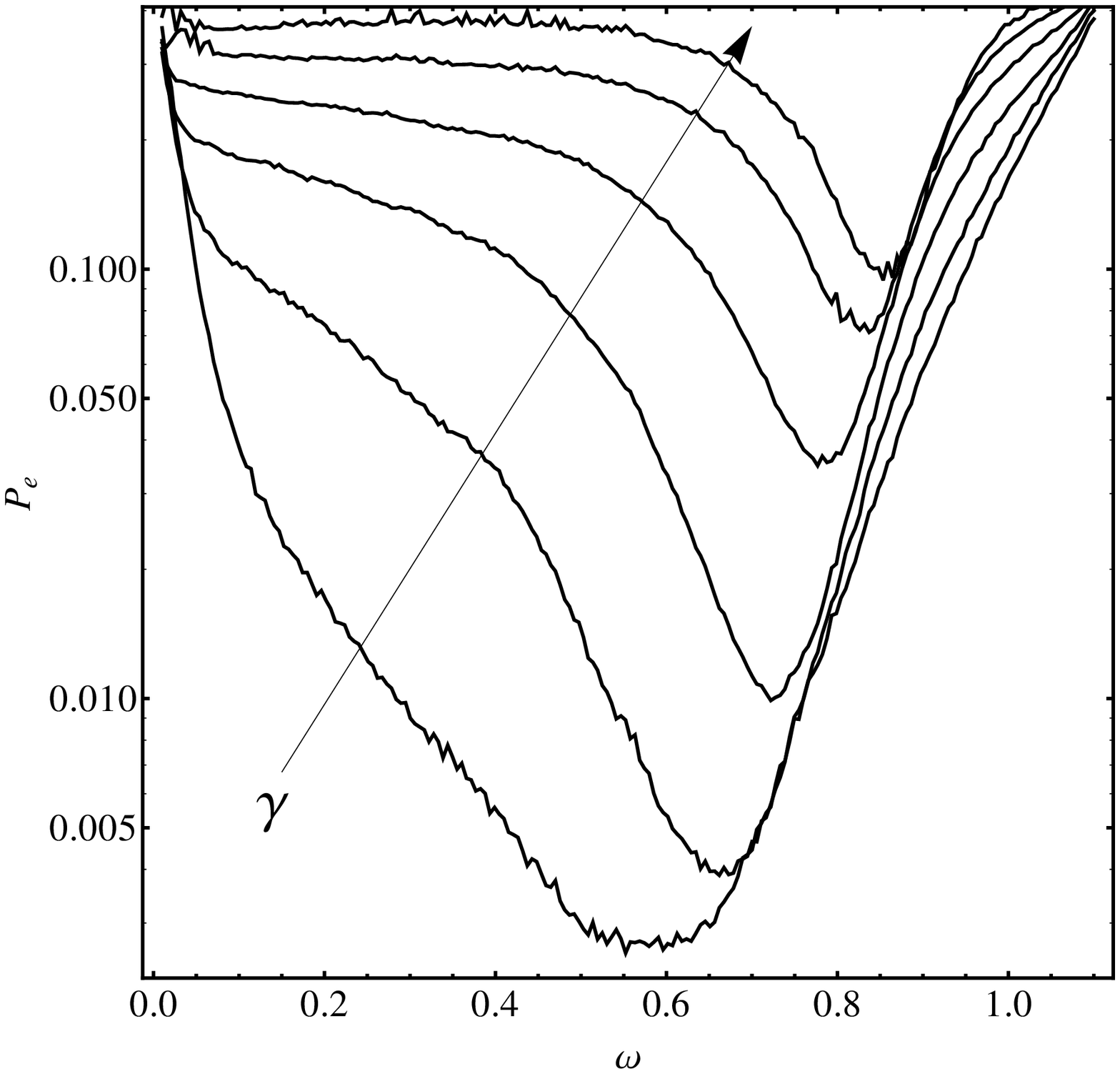}}
\caption{Multiple plot of the error probability $P_e$ as a function of the 
applied signal frequency $\omega$ and the bias current $\gamma$. (a): SM detector, (b) LRT detector.
Along the arrow bias $\gamma$ decreases from $0.9$ to $0.4$ with step $-0.1$.
Other relevant parameters of the system are: $\alpha = 0.05$, $\varepsilon_N = 0.0175$, $\varepsilon=0.05$ and $\varphi_0 = 0$. The simulations are performed setting 
the mean observation time under $\mathcal{H}_0$,  $E[T_{obs}|\mathcal{H}_0] =2000$.}
\label{fig:cont_omega_gamma}
\end{figure}

As shown in Sec. IV A the normalized frequency plays a major role in the performances, see Fig. \ref{fig:cont_omega_gamma}(a,b). A detector should therefore be optimized through an appropriate choice of the normalized applied frequency  $\omega$. This may be done tuning the Josephson angular velocity  close to the signal frequency. Therefore we conclude that signals in the range of $\omega_j$ are suitable for detection with JJ, while slower or faster signals are poorly analyzed with this technique.

\begin{figure}
\centerline{(a)}
\centerline{\includegraphics[scale=0.5]{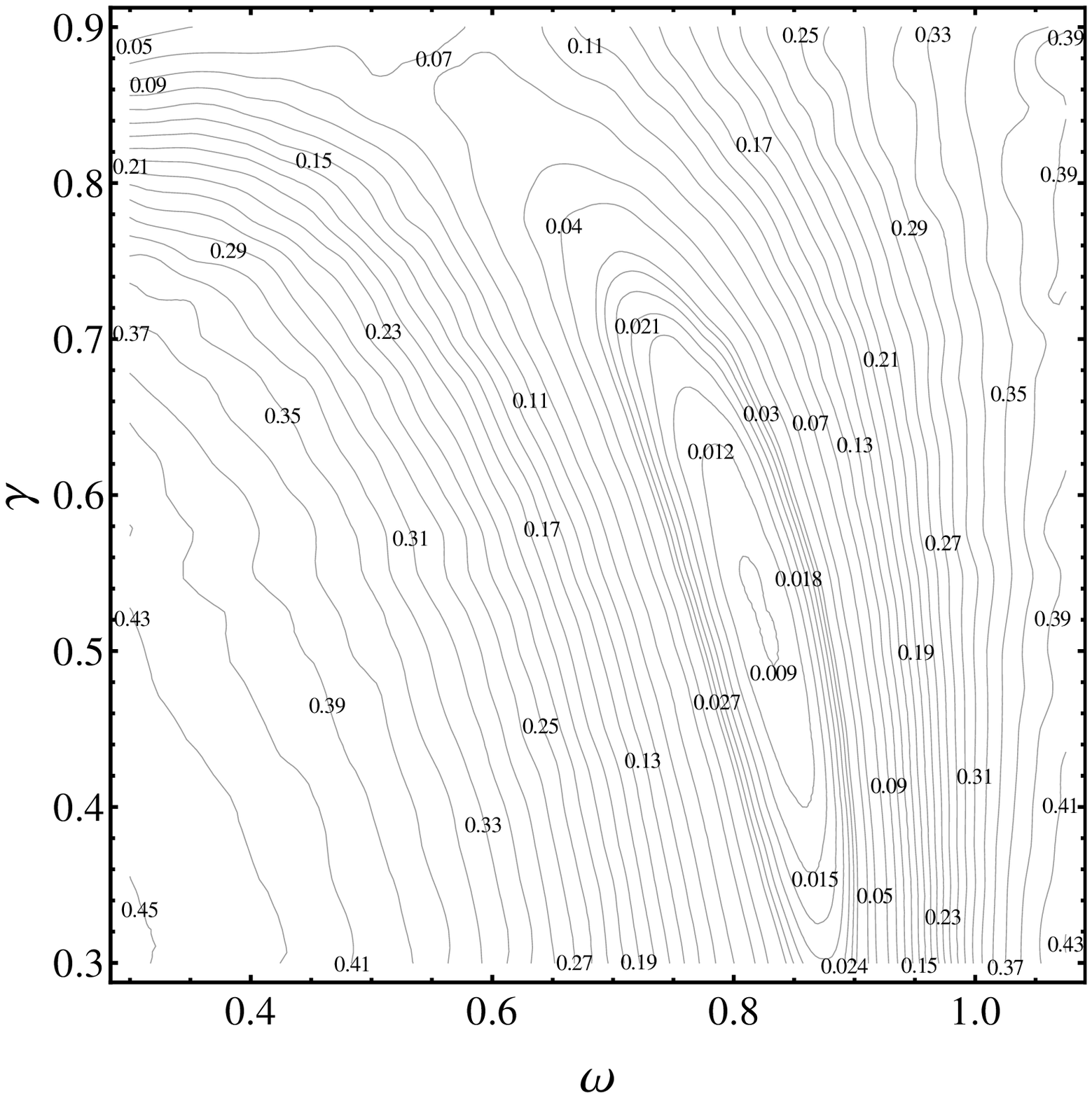}}
\centerline{(b)}
\centerline{\includegraphics[scale=0.5]{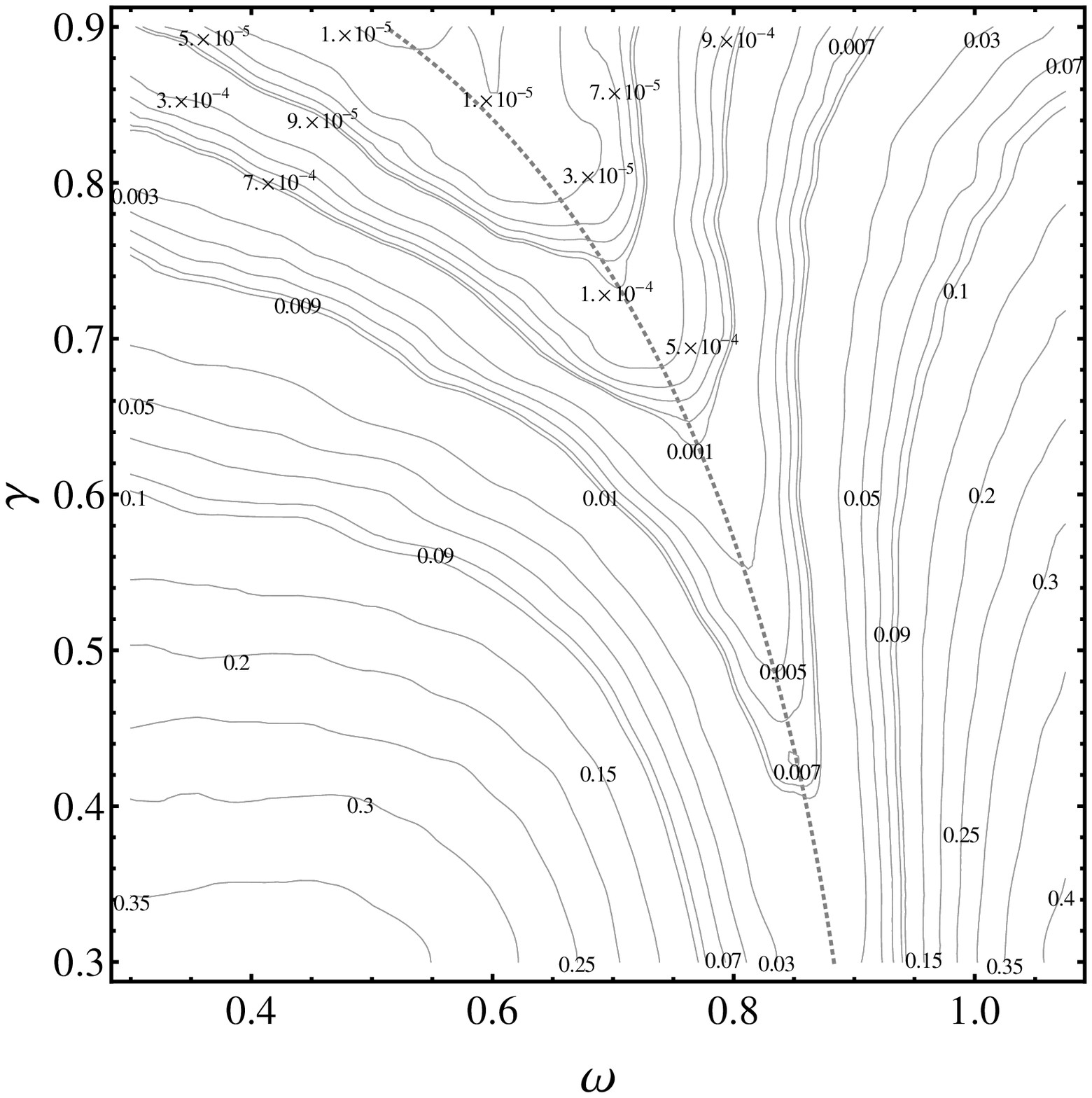}}
\caption{Contour plot of the error probability $P_e$ as a function of the applied signal frequency $\omega$ and the bias current $\gamma$. (a): SM detector, (b) LRT detector.
Thick line indicates in (b) the location of minimum $P_e$ probability. 
Other relevant parameters of the system are: $\alpha = 0.05$, $\varepsilon_N = 0.0175$, $\varepsilon=0.05$ and $\varphi_0 = 0$. The simulations are performed setting the 
mean observation time under $\mathcal{H}_0$, $E[T_{obs}|\mathcal{H}_0] =2000$.}
\label{fig:omega_gamma}
\end{figure}

Once the system is optimized in frequency, probably the most easily tunable parameter is the external bias current, $I_B$ that affects the normalized bias $\gamma$. In the geometric resonance neighborhood, as shown in Fig. \ref{fig:omega_gamma}(a), there is a clear optimal point around $\gamma \simeq 0.5$ for the SM technique. The physical interpretation is that the average escape time is most sensitive to the external signal when the bias current $\gamma$, and hence the energy barrier of Eq. (\ref{eq:barrier}), is intermediate between the maximum barrier ($\gamma=0$) and the minimum barrier ($\gamma=1$). 
Indeed in the former case there are few escape events in the observation time, while in the latter case the escape is dominated by noise and the average is little affected by the signal. It is interesting to see that for some high values of $\gamma$ the stochastic resonance dip around $\omega_{SR}$ performs better than the geometric resonance close to $\omega_{res}$. 
The LRT, as expected, performs better than SM for all $\gamma$ values, see Figs. \ref{fig:omega_gamma}(a,b). 
The LRT detector also shows a different behavior, namely the larger the bias current the more accurate the estimate, as displayed in Figs. \ref{fig:cont_omega_gamma}(b) and 
\ref{fig:omega_gamma}(b).  The qualitative explanation is as follows: with a more accurate analysis of the escape times it is possible to recognize the presence of the signal embedded in the noise even when the average is little affected by the signal itself, due to the optimal exploitation of the PDFs information - see  Eq. (\ref{eq:loglikeT}). 
It is important to notice that practical detectors cannot be realized just setting the bias current $\gamma=1$ for several reasons. 
First, the metastable state exists only for 
$\gamma < 1$. 
Second, in the analysis presented here we assume that the escape time is measured with an infinite accuracy. In practical detectors the finite error  associated to the measurements 
 is more relevant for shorter escape times.
We conclude that simulations suggest to use a bias current $I_B$ as close to $I_c$ as it is possible (i.e. $\gamma\simeq 1$), providing that the resulting escape times are still measurable with good accuracy. This entails that the limit of the performances is given by the actual electronics employed.

The critical current might be chosen to change the normalized drive amplitude $\varepsilon = S_0/I_c$ and the noise intensity $\varepsilon_N = \omega_jD/I^2_c $. As anticipated, the available range of the critical current is relatively narrow. However, we have found that an 
unifying parameter resumes the role of $\gamma$ (that depends upon the bias and can therefore be easily tuned) and $\varepsilon_N$.
Physical intuition suggests that escape time detector performances are mainly dependent on the potential barrier $\Delta U$ and the noise intensity $\varepsilon_N$. 
A realistic guess could be that a relevant parameter is the ratio $\rho$:

\beq
\rho = \frac{\Delta U(\gamma)}{\varepsilon_N} =\frac{2[\sqrt{1-\gamma^2} - \gamma \cos^{-1}(\gamma)]}{\varepsilon_N}.
\label{eq:rhobarratio}
\eeq

\begin{figure}
\centerline{(a)}
\centerline{\includegraphics[scale=0.5]{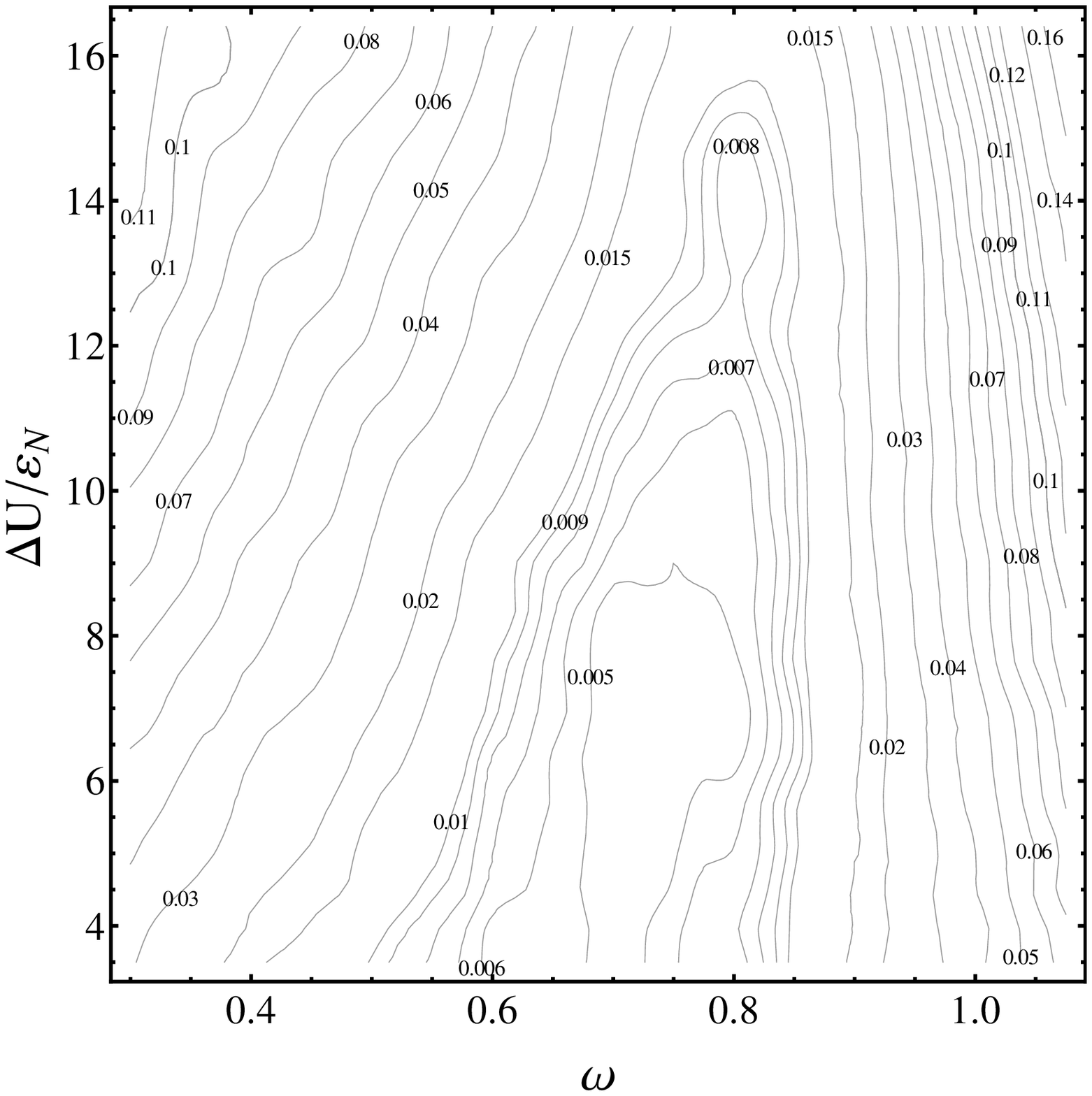}}
\centerline{(b)}
\centerline{\includegraphics[scale=0.5]{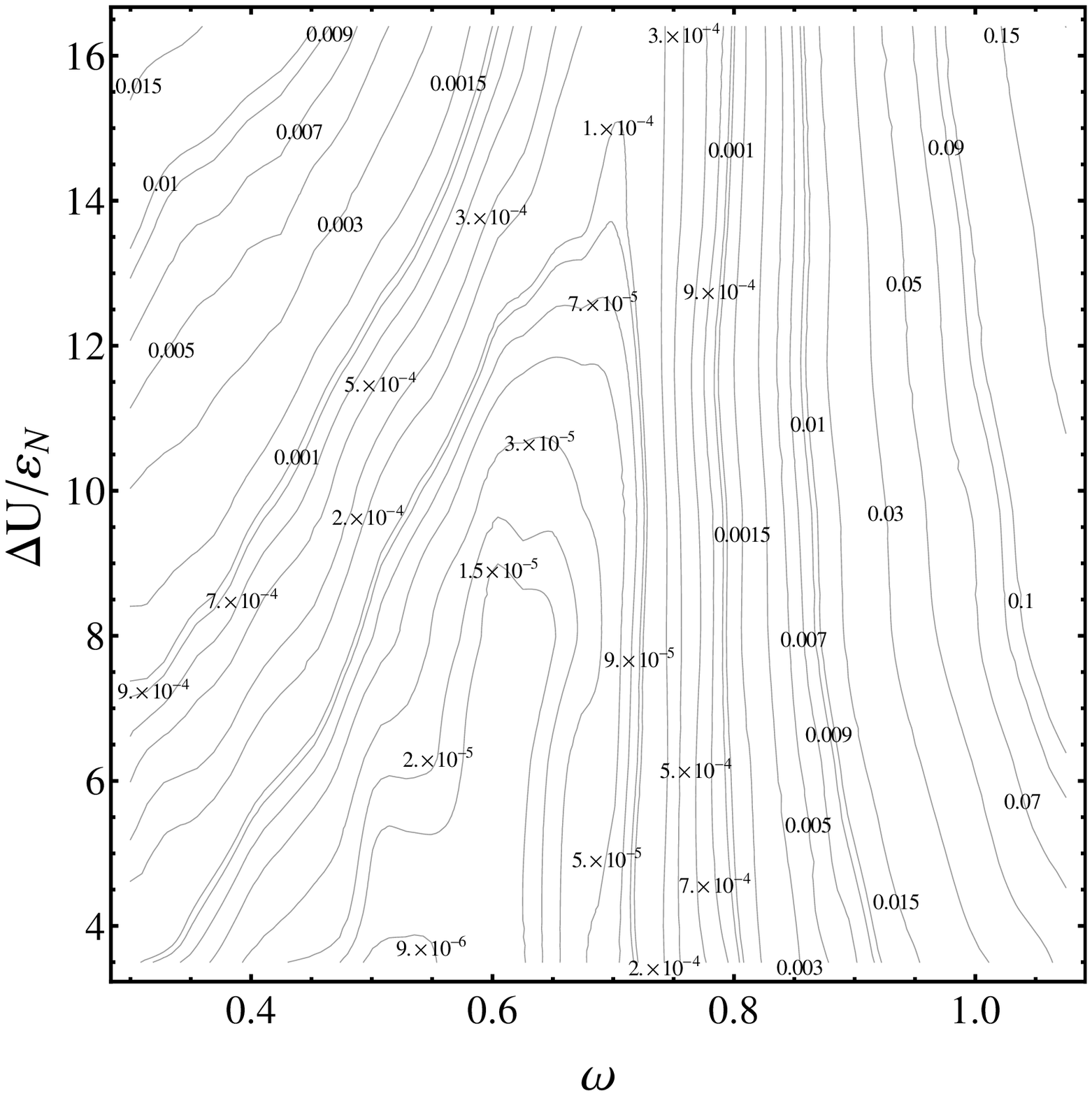}}
\caption{Contour plot of the error probability $P_e$ as a function of the applied signal frequency $\omega$ and the noise-normalized barrier $\Delta U/\varepsilon_N$. 
(a) $\varepsilon=0.1$ and  $\varepsilon_N=0.07$, (b) $\varepsilon=0.05$ and $\varepsilon_N=0.0175$. 
Other relevant parameters of the system are: $\alpha = 0.05$ and $\varphi_0 = 0$. The simulations are performed setting the mean 
observation time under $\mathcal{H}_0$, $E[T_{obs}|\mathcal{H}_0] =2000$.}
\label{fig:DUvsOmega}
\end{figure}

Simulations in Figs. \ref{fig:DUvsOmega} (that display 
the contour level of error probability as a function of $\Delta U/\varepsilon_N$ and $\omega$ for two different values of $\varepsilon$ keeping the same SNR $\propto \varepsilon/\sqrt{\varepsilon_N}$) confirm this conjecture. 
Despite the nonlinear character of the system, the two contour plots are qualitatively similar. The main difference is due to the rectification effect (see again Fig. \ref{fig:optres}). We conclude that the analysis of parameter $\rho$ leads to the same qualitative optimization recipes of parameter $\gamma$ : for the LRT strategy,  the normalized bias current 
should be close to unity, while for the SM strategy optimization requires an intermediate bias value.

Concerning dissipation,  simulations with different values $\alpha$ in the underdamped regime ($\alpha \le 0.1$) leads to the following result: 
lowering the dissipation the escape rate is increased, and therefore more events can be collected in the same time interval. 
As a consequence, the lower the dissipation the better the detector performances. So we conclude that one should try to use a junction resistance as large as possible to decrease dissipation.
Moreover, our findings (lower dissipation favors detection) indicate that an external shunt to decrease the junction resistance might result in worse performances.

Let us summarize the findings of the best JJ parameters for signal detection: 

\begin{enumerate}

\item The normalized frequency $\omega = \Omega/\omega_j$ shows best performances around the geometrical resonance of Eq. (\ref{eq:tau_barrier}), $\omega \simeq \omega_{res}$, for both LRT and SM. JJ are therefore best suited for signals whose frequency can be around $10-1000GHz$. If the simpler sample mean is employed, a second local optimum appears at a lower frequency, Eq. (\ref{eq:tau_barrier}), a frequency that can be also much smaller than the geometric resonance.

\item  The bias current for the LRT strategy should be set as close as possible to $\gamma = I_B/I_c=1$ to achieve the lowest value of $\Delta U$, see Fig. \ref{fig:cont_omega_gamma}(b). For the SM strategy the currents should be set at an intermediate value that depends on the ratio $\rho$ between the the energy barrier and the signal. 

\item The normalized signal $\varepsilon_N$ is not an independent variable, for the relevant parameter is the ratio between the normalized signal and the energy barrier,
 Eq. (\ref{eq:rhobarratio}). This ratio should be, for the LRT strategy, as small as possible, this is automatically set if the bias current condition has been fulfilled .

\item  The normalized dissipation parameter $\alpha$ 
should be as low as possible. A low dissipation in fact results in faster escapes, thus increasing the statistics at a fixed length of the signal. The other features (in terms of optimization of the normalized drive frequency, bias current, and signal amplitude) are independent of dissipation. We do not show the numerical results for different $\alpha$, for they are very similar to the displayed results obtained with $\alpha = 0.05$.
\end{enumerate}

The search for optimal parameters has led to qualitative indications (e.g., the indication for high bias and matching to the nonlinear resonance) 
to select the experimental set-up without actually repeating our numerical work. The purpose of the previous analysis is to determine the optimal working point location in the JJ parameter space at a fixed value of the SNR. On the other hand, SNR strongly affects the performance of the detector evaluated through the error probability.
Thus it is interesting to characterize the relationship between the SNR and the error probability $P_e$ for both SM and LRT detectors. 
Our \emph{ansatz} is that the performances measured by the KC index follow the power law
\beq 
d_{KC}(Y)\sim A(Y)\left(\frac{\varepsilon}{\sqrt{\varepsilon_N}}\right)^{\eta(Y)},
\label{eq:pap_dipende}
\eeq
where $Y$ is a general asymptotically Gaussian decision statistics. In our setup $Y$ is replaced by the average of the sampled escape times $\mathcal{A}(\underline{\tau})$ for the SM detector and by $\Lambda(\underline{\tau})$ for the LRT detector, defined in Eqs. (\ref{eq:smT}) and (\ref{eq:locloglik}), respectively.
Therefore, as described in Appendix B, the error probability can be expressed as
\beq
P_e=
\displaystyle{\frac{1}{2}\mbox{erfc}
\left( B(Y)
\left(\frac{\varepsilon}{\sqrt{\varepsilon_N}}\right)^{\eta(Y)}
\right)},
\label{eq:fitmode}
\eeq
The behavior for the optimal detection parameters has been verified via numerical simulation and the results are shown in Fig. \ref{fig:PevsSNR}(a).
It is evident that the model fit in Eq. (\ref{eq:fitmode}) is well suited for both SM and LRT detectors. 
More interesting is that the scaling law, i.e. the value of $\eta(Y)$, is different for the two detection strategies.
Indeed for the SM detector we have found $\eta(\mathcal{A}) \approx 1.63 \sim 3/2$ \cite{filatrella10}, while LRT shows a scaling law ruled by $\eta(\Lambda) \approx 0.97 \sim 1$, that is nearly optimal because it is also the behavior of the exponent for the ideal matched-filter (as can be promptly seen in Fig. \ref{fig:PevsSNR}(b)).
This striking difference between the SM and LRT is even more evident on the basis of the following consideration. The KC index is roughly proportional to $\sqrt{T_{obs}}$ via the coefficient $B(Y)$, hence lowering SNR the detection time should be extended to preserve the same detector performance. 
To keep constant the quality of the detection the observation time $T_{obs}$ should increase as $\left(\varepsilon/\sqrt{\varepsilon_N}\right)^{-3}$ for the SM detector and as
$\left(\varepsilon/\sqrt{\varepsilon_N}\right)^{-2}$ for the LRT one. Such a different scaling gives a huge advantage to the LRT in the challenging case of small SNR.
In Fig. \ref{fig:PevsSNR}(b) the comparison between the proposed LRT detection scheme and matched filter \cite{helstrom} is further elucidated by using a suitable
representation based on the KC index.
In this framework,  as anticipated, the $\eta(\Lambda)$ parameter  is readily seen  of the order of unity for all displayed curves.
Moreover it is simple to quantify the loss in decibel (dB) of the detector performance, by mean of the straight lines intercepts, that for 
the best case is nearly $4$ dB. Rephrasing this loss in terms of the ratio between LRT  and matched filter observation time (for continuous signals) corresponds to a factor $\sim 2.5$. 
Such speed up factor can be achieved by frequency scaling and/or parallel VLSI implementation
of the detector and JJ could turn out to be competitive in practical applications.     
In the inset of Fig. \ref{fig:PevsSNR}(b) it is shown the $d_{KC}$ loss with respect to matched filter in a decibel scale as a function of noise parameter $\varepsilon_N$.
The LRT, as expected, never passes the matched filter. The minimal loss is reached for
reasonably high values of $\varepsilon_N$ that corresponds to the physical limit of negligible escapes towards the higher local minima. Finally, we have verified that 
inside the optimality region this function is substantially
independent of the time window $T_{obs}$ and of the ratio $\varepsilon/\sqrt{\varepsilon_N}$.

\subsection{Unknown Phase Estimation and Signal Detection}

\begin{figure}
\centerline{\includegraphics[scale=.65]{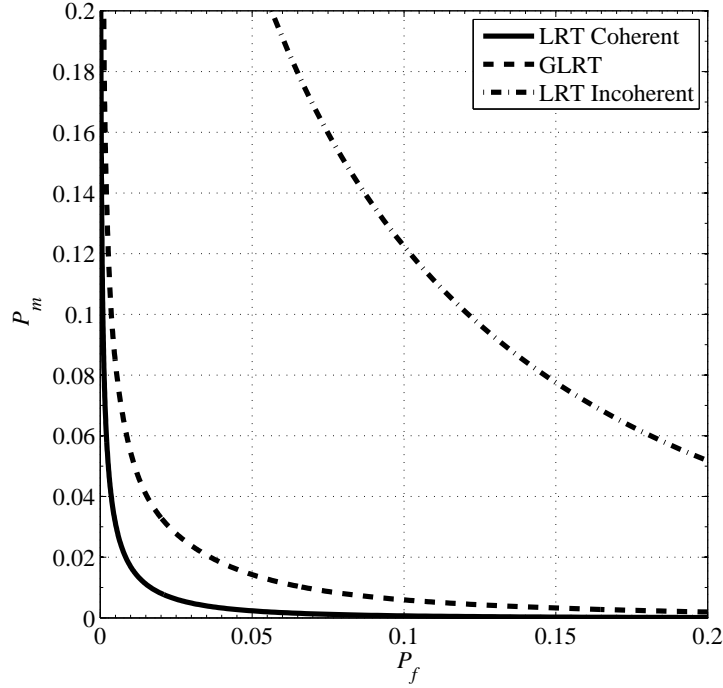}}
\caption{ROCs computed for three different detectors:
LRT detector based on incoherent acquisition of escape times (dot dashed line);  filter bank coherent acquisition (GLRT)
detector (dashed line) and  LRT detector based on coherent data (continuous line). For all strategies we ensure the same mean observation time under $\mathcal{H}_0$ (i.e. $E[T_{obs}|\mathcal{H}_0] \approx 1500$).
Other relevant parameters are:
$\gamma = 0.5$, $\alpha = 0.05$, $\varepsilon_N = 0.07$. Moreover, when the signal is present,  $\varepsilon = 0.1$, $\varphi_0 = 0$ and $\omega = 0.8$, close to the optimal detection point $\omega_{res}$.  }
\label{fig:coerVSincoer}
\end{figure}

\begin{figure}
\centerline{\includegraphics[scale=.5]{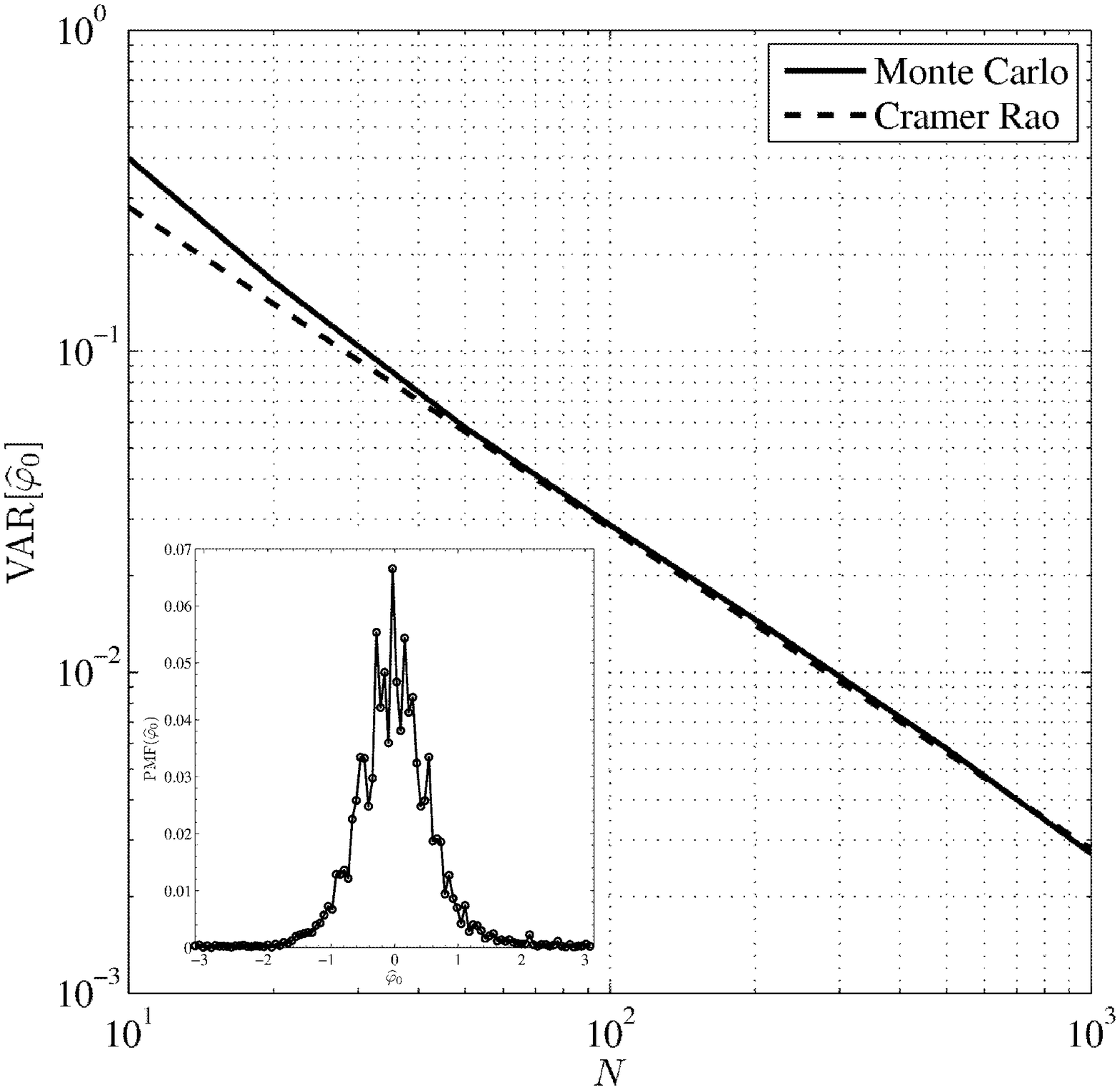}}
\caption{Variance of the maximum likelihood estimate of $\varphi_0$  as a function of the number $N$ of escape times.
Continuous line is obtained by Monte Carlo method, dashed line is the theoretical Cramer-Rao lower bound.
In the inset a typical probability mass function estimate (with $10^5$ trials) of the unknown parameter
 $\varphi_0$ by using only $N = 10$ samples. 
The signal is generated with an initial phase  $\varphi_0 = 0$. 
Other relevant parameters are:
$\gamma = 0.5$, $\alpha = 0.05$, $\varepsilon_N = 0.07$, $\varepsilon = 0.1$ and $\omega = 0.8$, close to the optimal detection point $\omega_{res}$. }
\label{fig:CramerRao}
\end{figure}

As anticipated  in Subsection III.C, incoherent strategy may be implemented with an acquisition method that requires the simpler measurement setup.
In Fig. \ref{fig:coerVSincoer} three different  ROC are displayed to compare the proposed detection strategies.
The figure shows the natural behavior, i.e. that the best detection performances are obtained with the LRT detector (solid curve). 
A filter bank (GLRT coherent strategy shown in dashed curve) results in an acceptable deterioration of the detector performance. 
As anticipated, due to the huge reduction of information on the initial phase the incoherent strategy outperforms the other strategies.
The dependence of the variance estimate $\varphi_0$ by the GLRT detector implemented via filter bank is illustrate in Fig. \ref{fig:CramerRao}.
The $M=100$ sampling points are placed in $\varphi_0^{(i)}=\frac{\pi}{M}(2i-1-M)$ in order to cover the interval $[-\pi, \pi]$.
The sinusoidal signal with known parameters $\varepsilon=0.1$, $\omega=0.8$ and $\varepsilon_N=0.07$, is applied to a JJ with $\alpha=0.05$ and $\gamma=0.5$.

The unknown parameter $\varphi_0=0$ is chosen in the middle position between two sampling points, the worst case for the filter bank performance \cite{Nota}. Fig. \ref{fig:coerVSincoer} also shows (continuous curve) that the variance reduces by increasing the number
$N$ of escape times (the acquisition is supposed coherent).
In particular the estimator variance reaches the Cramer-Rao lower bound relative to the escape time statistics, i.e. the asymptotic efficiency \cite{Shao}, for 
$N\sim 50$ escapes. The proposed estimator is therefore able to effectively extract the information about the initial signal phase $\varphi_0$ carried by the escape times. Of course, when for very large values of $N$ the expected variance becomes lower than the discretization error $(2 \pi/M)^2/12$, the latter dominates and causes a saturation of the performances.

\section{Conclusions}

We have investigated a case of detection of sinusoidal signals by means of a nonlinear device, namely an underdamped JJ. Some warnings are in order to clarify the limits of the approach. First, compared to traditional methods, JJ require cryogenic facilities to cool the superconducting electrodes. Second, the best performances are obtained at high frequency, thus requiring to handle microwaves without introducing distortions or undesirable noise increases. Third, we recall that a matched filter is the optimal method for signal detection, and therefore our proposal can be considered only when such optimal method is not available for technical difficulties. 
Forth, we underline that the improvements are obtained at the price of an heavier numerical work, since the escape time distribution is not theoretically known.

We have shown that the analysis of the escape times, an experimentally accessible quantity, can be performed through an intuitive approach, the estimate of the average, or with a more refined likelihood-based statistic.
The latter analysis leads to a significant improvement of the performances of the LRT with respect to SM, 
greatly reducing the error probability. The advantages of better detectors could result in solid improvements in terms of experimental time length and costs.

We have found that the conditions that favor detection (in terms of frequency, bias current, and dissipation) depend on the adopted  strategy (SM or LRT).
Moreover, we have found that the likelihood ratio test unexpectedly leads to an optimal scaling law for small SNR.
This result is surprising, because the
signal detection has been doubly deteriorated by the 
insertion of a non linear device and the change of the observable variable (i.e. escape time instead of the original signal noise mixture).
We have also shown that the LRT approach is scalable to detector arrays, thus allowing the signal phase estimate. However we believe that the most important finding is that, in some sense, the experimental setup should be tailored on the data analysis.
Indeed the two detection approaches can not be optimized with the same  normalized frequency and  applied bias current.
We speculate that the above results are fairly common to other activation detectors, because our analysis mostly relies on Arrhenius-like  activation law, that is independent on the energy barrier details. It is therefore possible that the advantages  can be exploited generalizing to other systems the methods we present in this work.

Future research will be devoted to the use of signal processing tools for thermal induced and quantum assisted transition rates discrimination \cite{Cirillo}.
This difference is particularly difficult to detect because microwave source in JJ in Macroscopic Quantum Tunneling regime
produces multiple photon quantum transitions
that can be confused with subharmonic excitations due to nonlinear
JJ behavior \cite{Rotoli06}.

\section*{Acknowledgments}
We acknowledge Giacomo Rotoli for a careful reading of the manuscript and  Sergio Pagano for fruitful discussions.
 This work has been supported by the Italian Super Computing Resource Allocation 
ISCRA, CINECA, Italy (Grant IscrB\_NDJJBS 2011).

\appendix

\section{Kernel Density Estimation.}

The Likelihood Ratio Test, presented in Eq. (\ref{eq:likeT}), requires the complete knowledge of the PDFs $f_0(\cdot)$ and $f_1(\cdot)$ for both the hypothesis. 
As already mentioned, unfortunately there is no theoretical result that can provide this knowledge. Thus we are compelled to use an estimated version of these PDFs. 
A simple but effective strategy is to use the Kernel Density Estimation \cite{silverman}. This technique generalizes the basic idea of histogram by using
 a so-called \emph{Kernel function} $K(\cdot)$ that usually is a symmetric PDF. 
If there is a random sample $\underline{X} = \left\{X_i\right\}_{i\in[1, N]}$, where $N$ is the sample size, the kernel estimator should be 
\beq
\widehat{g}(x) = \frac{1}{N w} \sum_{i=1}^N K \left(\frac{x-X_i}{w} \right).
\label{eq:KDENaive}
\eeq
where the parameter $w$ is the \emph{bandwidth} (also called \emph{smoothing parameter}).
If we apply this framework to escape times, we immediately encounter a first difficulty. Indeed escape times are by definition \emph{positive}, i.e. their PDF,
 under the generic hypothesis $\mathcal{H}_j, ~j\in\{0,1\}$, has the property
\beq
f_{j}(t) = 0, ~~\forall t < 0.
\label{eq:posdef}
\eeq
Equation (\ref{eq:KDENaive}) leads to an estimated PDF that does not satisfy the inequality (\ref{eq:posdef}).
To deal with this issue, the following procedure has been applied.
\begin{itemize}
\item[a)] For a fixed hypothesis $\mathcal{H}_j$, the random sample of escape times 
$\underline{\tau} = \left\{\tau_i\right\}_{i\in[1, N]}$ is transformed via 
\beq
X = \log(\tau).
\label{eq:rvtr}
\eeq
Thus we deal with the random sample $\underline{X} = \left\{X_i\right\}_{i\in[1, N]}$ that can assume every value on the real axis.
\item[b)] By means of Eq. (\ref{eq:KDENaive}), an estimated PDF $\widehat{g}_{j}(x)$ is computed. 
\item[c)] The PDF $\widehat{f}_{j}(t)$ is obtained from $\widehat{g}_{j}(x)$ via 
\beq
\widehat{f}_{j}(t) = \frac{\widehat{g}_{j}(\log(t))}{t}, ~~t>0.
\eeq
\end{itemize}
The procedure described above has to be applied twice, i.e. under both the hypothesis $\mathcal{H}_j$, 
to obtain the estimates  $\widehat{f}_{0}(\cdot)$ and $\widehat{f}_{1}(\cdot)$ PDF. 
Then, when we have to decide about the presence of a signal via an independently generated random sample of escape times,
we can compute the statistic in Eq. (\ref{eq:loglikeT}) as:
\beq
\widehat{\Lambda}(\underline{\tau}) = \frac{1}{N}\displaystyle{\sum_{i = 1}^{N}\log 
\left[ \frac{\widehat{g}_{1}(\log(\tau_i))}{\widehat{g}_{0}(\log(\tau_i))} \right]}
\begin{array}[c]{c}
{\mathcal{H}_1}\\{>} \\
{<}\\{\mathcal{H}_0} 
\end{array}
\zeta,
\label{eq:estloglikeT}
\eeq
The procedure is completely specified if in Eq. (\ref{eq:KDENaive}) one chooses both the kernel function and the bandwidth. We have used the kernel defined by a standard Gaussian density, i.e. 
\[
K(t) = \frac{1}{\sqrt{2 \pi}} ~\displaystyle{\exp\left(-\frac{t^2}{2}\right)}.
\]
This kernel is rewarding, for the estimated density is smooth and there are no subset on the real axis with zero density. The latter point is essential in the likelihood ratios to avoid singularities.
The drawback is in a little loss in estimation performances, that is negligible for the used sample size ($\sim 5\cdot10^5$). 
The smoothing parameter selection, instead is an hard task. Indeed it is known (see \cite{silverman}) that the optimal choice has the form
\beq
w = C(K,f) N^{-1/5},
\eeq
where $C(K,f)$ is a constant that depends on the used kernel $K(\cdot)$ and, unfortunately, on the same density $f(\cdot)$ to be estimated.
Thus we have used a simple closed form obtained when both $K(\cdot)$ and $f(\cdot)$ are Gaussian that is
\beq
w = \left(\frac{4}{3}\right)^{1/5} \widehat{\sigma} N^{-1/5},
\eeq
where $\widehat{\sigma}$ is the sample standard deviation of the dataset.
This choice is near-optimal for unimodal densities, that is the case when the $\mathcal{H}_0$ hypothesis is in force, while 
it leads to some over-smooth for multimodal densities that arise under $\mathcal{H}_1$ hypothesis. Also in this case the large sample used helps us make the over smoothing negligible. Moreover, the slight over-smoothing introduced under $\mathcal{H}_1$ can only worsen the LRT test performance, for the PDF oscillations, that contain the largest part of additional information with respect to the sample mean, are underestimated. 


\section{Kumar-Carrol index.}
\begin{figure}
\centerline{(a)}
\centerline{\includegraphics[scale=0.5]{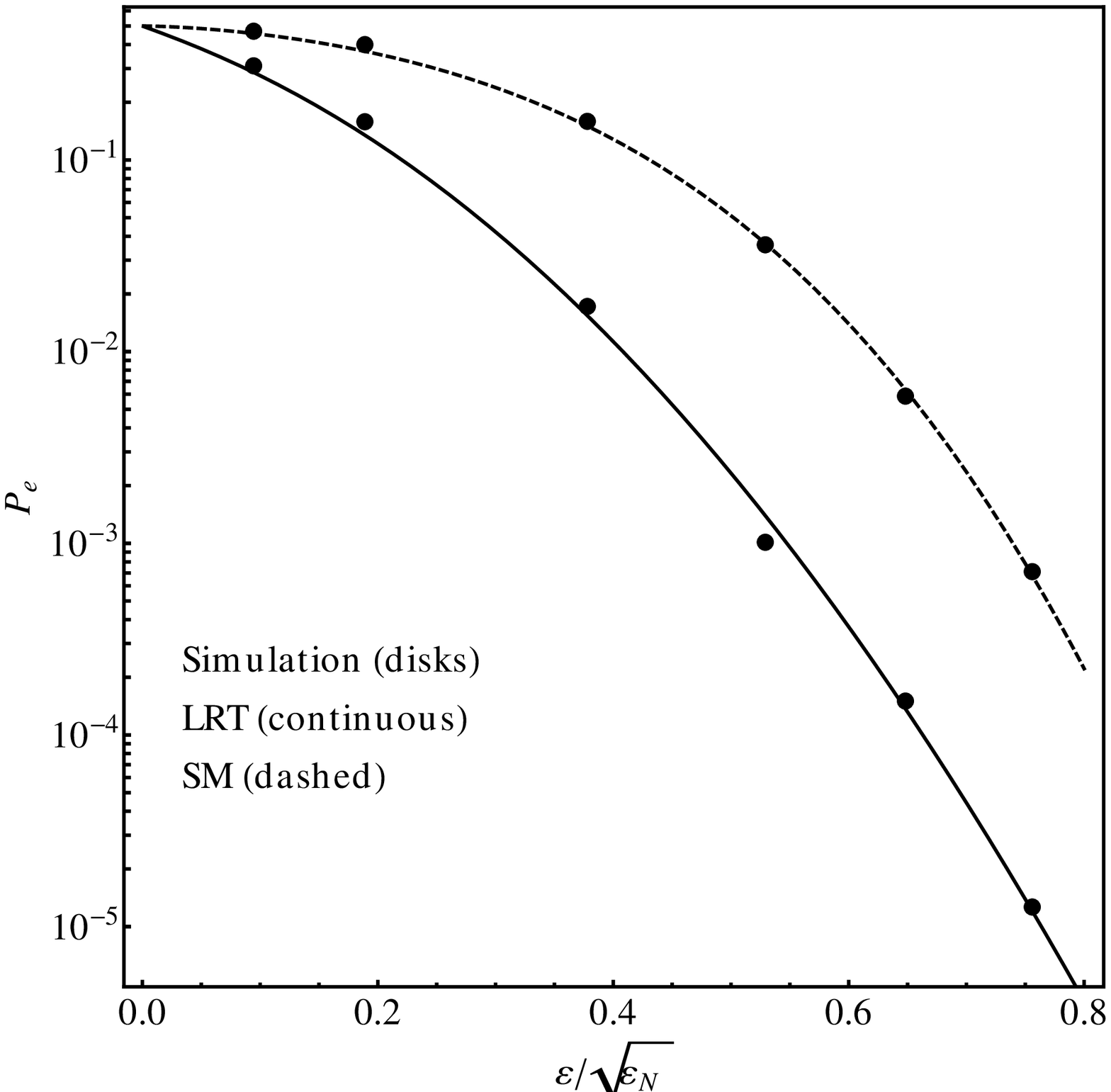}}
\centerline{(b)}
\centerline{\includegraphics[scale=0.33]{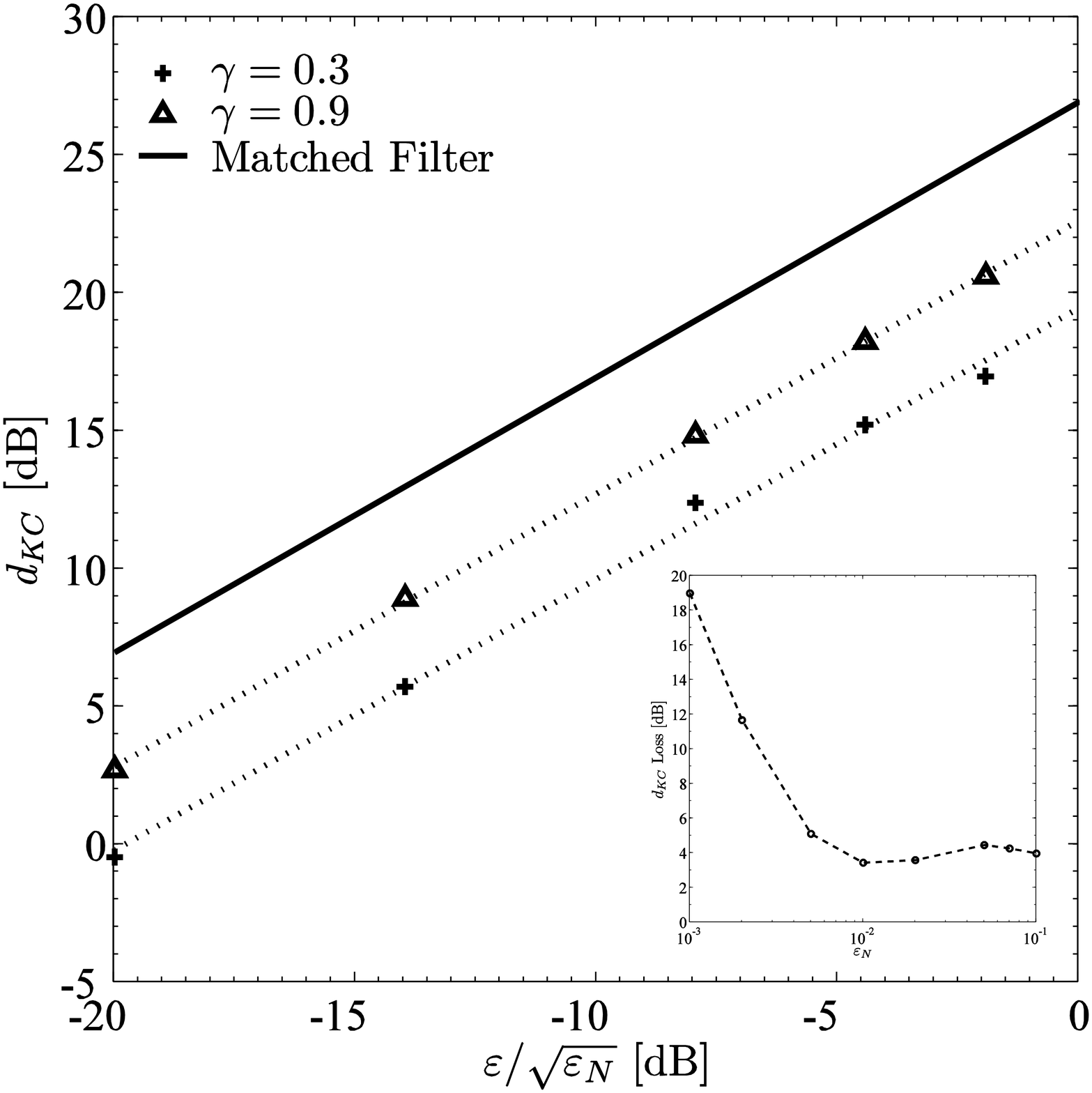}}
\caption{In the part (a) we display the error probability $P_e$ as a function of $\varepsilon/\sqrt{\varepsilon_N}$ (related to SNR).
Disks are obtained by numerical simulations, continuous lines are computed by fitting the data using the KC approximation in Eq. (\ref{eq:fitmode}). 
 The parameters used in the simulations are:
$\gamma = 0.3$, $\alpha = 0.05$, $\varepsilon_N = 0.07$. 
Moreover, when the signal is present, $\varphi_0 = 0$ and $\omega = 0.8$, close to the optimal detection point $\omega_{res}$.
Below in (b) the KC index (reported in dB) of proposed LRT strategy (for different parameters $\gamma=0.3, 0.9$ and $\omega$ near to the optimal detection frequency) 
vs. the same SNR related parameter is compared with the KC index of the
Matched Filter (best available strategy). In the inset the $d_{KC}$ loss w.r.t. matched filter (in a dB scale) for $\gamma=0.9$ is displayed as a function of $\varepsilon_N$
($\varepsilon/\sqrt{\varepsilon_N}$ is fixed to $-20$ dB). In all cases simulations are performed with $E[T_{obs}|\mathcal{H}_0] =2000$.}
\label{fig:PevsSNR}
\end{figure}

In this Appendix we introduce the general concept of KC index and its connection with a more
effective parameter: the  error probability.
In the contest of decision theory , the main purpose is to discriminate  between two hypothesis: signal is present ($\mathcal{ H}_1$) vs signal is absent  ($\mathcal{ H}_0$). Let $Y$ the selected decision statistics, expressed as function of the escape times vector $\underline{\tau}$, that has to be compared with a suitable threshold $\zeta$. 
Moreover, suppose that 
 $\mu_1(Y)$ and $\mu_0(Y)$ 
are the averages of the decision statistic under the two hypothesis (subscripts $1$ and $0$ refer to the presence and absence of the signal, respectively)
while the corresponding standard deviations are denoted with  $\sigma_1(Y)$,$\sigma_0(Y)$. 
The KC index can be accordingly defined as
\beq
d_{KC}(Y)= \frac{\mid \mu_1(Y)-\mu_0(Y)\mid}{
\sqrt{ \frac{1}{2}\left( \sigma_{1}^2(Y)+\sigma_{0}^2(Y) \right)}} .
\label{appendix_KC}
\eeq
In this paper both the decision statistics can be modelled as the sample mean of a suitable random variable, i.e. $\tau$ for SM detector and $\mathcal{L}$ for LRT one (see Eqs. (\ref{eq:smT}) and (\ref{eq:loglikeT}) respectively).
In this case both the statistics, for large sample size, are asymptotically normal due to the Central
Limit Theorem \cite{Billingsley}.
Therefore the error probability  $P_e$ (defined above as  the value of the ROC  $P_{m}=P_{f}$) can be expressed as
\beq
P_e=
\displaystyle{\frac{1}{2}\mbox{erfc}
\left(\sqrt{1+\frac{\Delta(Y)^2}{4}} 
\frac{d_{KC}(Y)}{2\sqrt{2}}
\right)}
\label{eq:perrvsdkc}
\eeq
where $\Delta(Y)=2 \mid \sigma_0(Y)-\sigma_1(Y) \mid/ \mid \sigma_0(Y)+\sigma_1(Y) \mid$.
By inspection it can be shown that Eq. (\ref{eq:perrvsdkc}) is a decreasing function of $\Delta$, and therefore neglecting the difference among standard deviations (if it exists) it is possible to retrieve an upper bound of $P_e$ that is only function of $d_{KC}$, i.e.
\beq
P_e\le 
\displaystyle{\frac{1}{2}\mbox{erfc}
\left( 
\frac{d_{KC}(Y)}{2\sqrt{2}}
\right)}.
\label{eq:inequa}
\eeq
The inequality (\ref{eq:inequa}) clarifies the heuristic character of the KC index as an indicator of the detector performance.
The concept above introduced can be applied to SM and LRT detectors, by setting $Y := \mathcal{A}(\underline{\tau})$ and $Y := \Lambda(\underline{\tau})$, respectively.
Under the hypothesis that a power law relationship between KC index and
SNR exists, as exemplified by 
\beq 
d_{KC}(Y)\sim A(Y)\left(\frac{\varepsilon}{\sqrt{\varepsilon_N}}\right)^{\eta(Y)}.
\label{eq:dipende}
\eeq
The error probability can be expressed as
\beq
P_e=
\displaystyle{\frac{1}{2}\mbox{erfc}
\left( B(Y)
\left(\frac{\varepsilon}{\sqrt{\varepsilon_N}}\right)^{\eta(Y)}
\right)},
\label{eq:fitmode_App}
\eeq
where $B(Y)=\sqrt{1+\frac{\Delta(Y)^2}{4}}A(Y)/(2\sqrt{2})$.
This latter equation has been used to interpret the relationship between $P_e$ and $\varepsilon/\sqrt{\varepsilon_N}$ in Sec. IV, see Fig. \ref{fig:PevsSNR}.

\newpage

\newpage

\newpage

\newpage

\newpage

\newpage

\newpage

\newpage

\begin{thebibliography}{99}
\bibitem{Nota}{ We recall that although the phase is unknown, in the coherent strategy each detector is strobed by using the signal period. 
This ensures that we keep the initial (unknown) phase constant. Had we chosen the incoherent strategy, we would have lost all information about the phase.}
\end{thebibliography}
\end{document}